\pdfoutput=1
\RequirePackage{fix-cm} \documentclass[reprint,aps,prfluids,showpacs,superscriptaddress,amsmath,onecolumn,longbibliography,12pt]{revtex4-2}

\usepackage{graphicx}
\usepackage{dcolumn}
\usepackage{bm}
\usepackage[utf8]{inputenc}
\usepackage[T1]{fontenc}
\usepackage{mathptmx}
\usepackage{physics}
\usepackage{subcaption}
\usepackage{csquotes}

\graphicspath{{.}{figs/}{figs/logos/}}
\newcommand{\freeenergy}{\mathcal{F}}

\newcommand{\eps}[1]{\epsilon_{\text{#1}}}
\newcommand{\sig}[1]{\sigma_{\text{#1}}}

\begin{document}
\preprint{}

\title[Multiscale perspective on wetting on switchable substrates]{Multiscale perspective on wetting on switchable substrates: mapping between microscopic and mesoscopic models}
\author{Moritz Stieneker}
 \affiliation{University of Münster, Institute for Theoretical Physics, Wilhelm-Klemm-Str. 9, 48149 Münster, Germany}\affiliation{University of Münster, Center of Nonlinear Science (CeNoS), Corrensstr. 2, 48149 Münster, Germany}\author{Leon Topp}\thanks{M.\,Stieneker and L.\,Topp contributed equally to this work.}
\affiliation{University of Münster, Institute for Physical Chemistry, Correnstr. 28/30, 48149 Münster, Germany}

\author{Svetlana Gurevich}
\email{gurevics@uni-muenster.de}
 \affiliation{University of Münster, Institute for Theoretical Physics, Wilhelm-Klemm-Str. 9, 48149 Münster, Germany}\affiliation{University of Münster, Center of Nonlinear Science (CeNoS), Corrensstr. 2, 48149 Münster, Germany}

\author{Andreas Heuer}
\email{andheuer@uni-muenster.de}
\affiliation{University of Münster, Institute for Physical Chemistry, Correnstr. 28/30, 48149 Münster, Germany}\affiliation{Center of Nonlinear Science (CeNoS), University of Münster, Corrensstr. 2, 48149 Münster, Germany}
\affiliation{University of Münster, Center for Multiscale Theory and Computation (CMTC), Corrensstr. 40, 48149 Münster, Germany}

\date{\today}

\begin{abstract}
To understand the non-equilibrium relaxation dynamics of a liquid droplet on a switchable substrate the interplay of different length- and
time-scales needs to be understood.  We present a method to map the microscopic information, resulting from a molecular dynamics simulation, to a mesoscopic scale, reflected by a thin film model. After a discussion of the mapping procedure we first analyze the relaxation of a liquid droplet upon switching the wettability of the substrate. Further, we show that a nearly identical mapping procedure can be used for the description of two coalescing droplets. With our procedure we take a first step to extend the mapping from the equilibrium case to non-equilibrium wetting dynamics, thus allowing for a quantitative multi-scale analysis.

 \end{abstract}

\maketitle

\section{Introduction}

Finding ways to manipulate and control patterns of liquids has been the goal of scientists for a long time. Back in 1992, \citeauthor{ChWh1992s} were able to exploit a wettability gradient to make a droplet walk up an incline~\cite{ChWh1992s}. Another example includes extensively experimental studies of the instabilities, dynamics, and morphological transitions of patterns in thin liquid films on different pre-structured substrates, see e.g.,~\cite{GHLL-Science-99, B800121A,Seemann1848,Wang-Small-11,ZW-AMI-16}. Nowadays such systems can also be examined theoretically on different length and time scales~\cite{KS-PRL01,KLRD-PF-06,wilczek2016dip,Wang-NJP-2016,HLHT2015w,LeLi1998phys.rev.lett.}, e.\,g. transversal instabilities of ridges on pre-structured substrates have been studied with a combination of a microscopic kinetic Monte Carlo model and a continuum thin film model~\cite{TBHT2017tjocp,Wu2014,ZHANG2017886}.

In recent years the development of switchable surfaces gained pace. On such surfaces the wettability can be varied by applying an external stimulus like a change of the pH value or by illumination with light of a defined wavelength. Prominent examples for such surfaces are inorganic materials like TiO$_{2}$ or ZnO~\cite{Wang1997, tio22005, Sun2001} which have the advantage of a large difference between the contact angles before and after switching. Since the switching process from the hydrophilic to the hydrophobic case for these substrates is rather slow, another class of substrates is of great interest, namely substrates coated with a self-assembled monolayer (SAM) consisting of molecules with azobenzene or other photoresponsive moieties~\cite{Ishihara1982, Rosario2002,D0CC00519C}. The azobenzene moiety can be switched with UV light from a trans to a cis state which has a lower wettability while the reverse process can be induced by illumination with blue light. These surfaces adapt much faster at the disadvantage of lower contact angle differences. However, in recent years improvements which yield a higher change of the contact angle have been made by microstructuring the surface~\cite{B504479K, ruehe2012}.

Switchable substrates promise rich non-equilibrium behavior and an additional mechanism to control pattern formation, which can be employed in addition to static pre-structures. In particular it was demonstrated that it is possible to guide the movement of a droplet in a reversible manner by applying a light gradient~\cite{Ichimura1624, D0CC00519C}, i.\,e. changing the wettability close to the droplet. Recent theoretical work by \citeauthor{GrSt2021softmatter}~\cite{GrSt2021softmatter} investigated how droplets can be steered with the help of spatio-temporal wettability patterns using the macroscopic boundary element method. This is relevant especially for the development of lab-on-a-chip devices~\cite{SqQu2005rev.mod.phys.}.

Theoretical models play a key role to gain an improved understanding of the non-equilibrium behavior on switchable substrates. In particular, to study microscopic phenomena atomistic simulation methods like Molecular Dynamics (MD) have become an established approach~\cite{Allen2017,ruijter1999, ROYCHOUDHURI2020100628}. On larger length and time scales mesoscopic thin film descriptions have been successfully applied for a variety of different wetting systems, see the reviews~\cite{OrDB1997rmp,BEIM2009romp,Thie2010}. While microscopic MD simulations can incorporate more details of the specific interactions between liquid and substrate, continuum mesoscopic models cannot resolve microscopic details but are able to address much larger length and time scales. Furthermore, continuous mesoscopic models allow to apply the tool kit of bifurcation analysis to investigate instabilities offering analytical insights which is not possible in discrete, microscopic models. Bifurcation analysis combined with parameter sweeps, which are computationally cheaper compared to microscopic models, can then indicate interesting parameter regimes and time scales to analyze in the microscopic model for a more detailed investigation. This helps avoiding computational costs for simulations in irrelevant regimes.

Thus, combining different microscopic and continuum descriptions seems natural and has been done by \citeauthor{Wu2014}~\cite{Wu2014}, among others. There, spreading dynamics of drops on solid surfaces was investigated by solving the Navier–Stokes equations in a continuum domain comprised of the main body of the drop together with MD simulations in a particle domain in the vicinity of the contact line. Another example for the combination of models across length and time scales is the work by \citeauthor{ZHANG2017886} who combined MD with volume of fluid simulations to study droplet spreading on surfaces~\cite{ZHANG2017886}. Also \citeauthor{Hadjiconstantinou1999} supplied both, a continuum and MD method for the flow of two immiscible fluids in a channel~\cite{Hadjiconstantinou1999}. In~\cite{TBHT2017tjocp} kinetic Monte Carlo simulations and a thin-film continuum model were combined to comparatively study the Plateau-Rayleigh instability of ridges formed on pre-structured substrates. It was shown that the evolution of the occurring instability qualitatively agrees between the two models. 

Given the advantages and disadvantages of the different  methods, it is evident that a mapping between the methods is of great interest.
For the static case various microscopic descriptions have been employed to improve the mesoscopic models mainly by extracting the binding potential (also referred to as the wetting, disjoining or interface potential)~\cite{deGennes-RMP-85,RefJ_review}. For partially wetting liquids, the interface potential is particularly important for describing the droplets in the vicinity of the three-phase contact line. It is defined for a uniform thickness layer of the liquid on a flat solid wall in the presence of a bulk vapor phase. 

In particular, \citeauthor{TMTT2013tjocpb} extracted properties from a MD model to study equilibrium properties in a continuum model and found quantitative agreements between the MD model and the continuum model~\cite{TMTT2013tjocpb}. Similar results can be obtained based on density-functional theory~\cite{RefI_review, RefJ_review}. The results obtained in~\cite{RefI_review,RefJ_review} could be verified by a different method of extracting the disjoining pressure, namely using nudged elastic band calculations~\cite{RefK_review}. \citeauthor{RefJ_review} could relate oscillatory disjoining pressures to layering effects and found qualitative agreements to profile shapes observed in experiments~\cite{RefJ_review}. 

However, so far the focus has been on static equilibrium conditions. As switchable substrates inherently lead to non-equilibrium dynamics, static considerations are not sufficient anymore. Therefore, our focus here is on the mapping of the time scales.

In this paper we propose a general method to map a MD model to a mesoscopic thin film model making the first step towards quantitative comparisons of dynamics between models acting on mesoscopic and microscopic scales.

The functional form of the presented mapping can help to understand differences between the employed models and shed light upon the corresponding time scales and transport quantities. A mapping between the model parameters can possibly provide insights, if continuum models are able to grasp all the features present in particle-based, microscopic models. 

The paper is structured as follows: In section~\ref{sec:theory} we describe our simulation setups for MD and TFE simulations. Afterwards in section~\ref{ssec:how_to_compare_scales} our procedure for the spatial and temporal mapping between MD and TF simulations is introduced. Two applications - a droplet adapting to a new wettability and the coalescence of two droplets - are shown in section~\ref{sec:switching} and \ref{sec:coalescence_results}. Finally, we conclude our results in section \ref{sec:conclusion}

\section{Theoretical Background} \label{sec:theory}
\subsection{Molecular Dynamics theory}
\begin{figure}
\includegraphics[width=\columnwidth]{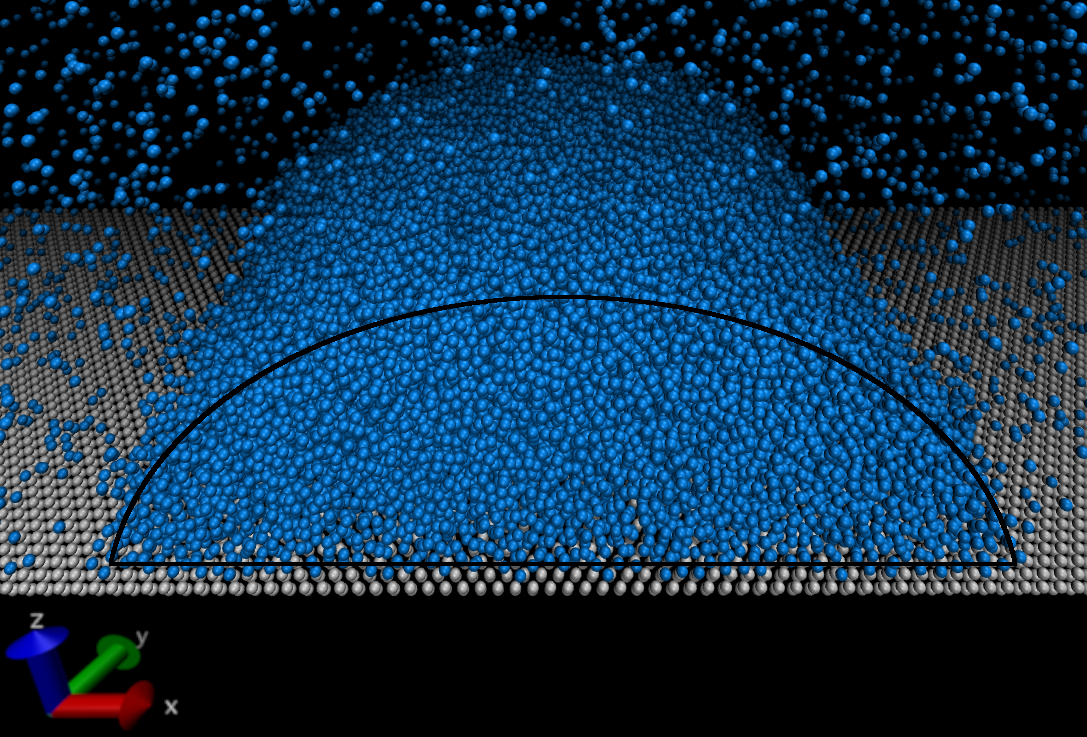}
\caption{Snapshot from a MD simulation showing fluid particles (blue) forming a cylindrically shaped droplet placed on two layers of a fcc(111) surface.}
\label{fig:snapshotMD}
\end{figure}
On the microscopic level, we employ the framework HOOMD~\cite{anderson_hoomd} to perform simulations of a system consisting of Lennard-Jones particles in the canonical ensemble.
All particles in the system are interacting through the Lennard-Jones potential
\begin{equation}
  V(r_{lj}) = 4 \eps{lj} \left( \left(\frac{\sig{lj}}{r_{\text{lj}}}\right)^{12} - \left( \frac{\sig{lj}}{r_{\text{lj}}} \right)^{6} \right),
\end{equation}
where $r_{\text{lj}}$ is the distance of particles $\text{l}$ and $\text{j}$, $\eps{lj}$ the interaction strength between
the particles and $\sig{lj} = \frac{1}{2} (\sig{l} + \sig{j})$ the mean of the particles' diameters $\sig{l}$ and $\sig{j}$.
This potential is truncated and shifted at a cut-off radius of $r_{c} = 2.5 \sigma$. 
The interaction strength is calculated as the geometric mean of the self interaction parameters $\eps{lj} = \sqrt{\eps{l} \eps{j}}$.
We distinguish two types of particles, namely the substrate particles (here denoted with "$s$"), which are fixed at their positions during the simulation, and the fluid ones (denoted with "$f$") which form
the droplet or are in the gas phase. While the interaction strength of the droplet particles is set to $\eps{f} = 1$ for all
simulations, by varying the parameter $\eps{s}$ the wettability of the surface can be changed. Note that $\eps{w}= \sqrt{\eps{s}\eps{f}}=\sqrt{\eps{s}}$ denotes the interaction strength between a fluid and substrate particle, because we set $\eps{f}=1$. In all simulations we vary the value $\eps{w}$ between $\eps{HW} = 0.762$ for a high wettability and $\eps{LW} = 0.632$ for a low wettability if not denoted otherwise. The particle diameter is set to $\sig{f} = \sig{s} = \sigma$ for the solid as well as for the fluid particles. The time step is set to $\tau = \sigma^{-1}\sqrt{\epsilon_{f}/M}/200$, where $M$ is the particle mass and the reduced temperature is $\frac{k_{B}T}{\epsilon_f}=0.75$.
The substrate particles are arrayed in two layers of a fcc(111) lattice. To exclude the effect of line tension we consider cylindrically shaped droplets as can be seen in Fig.\,\ref{fig:snapshotMD}. In $y$-direction our domain is 48.6~$\sigma$ wide. We performed simulations with an approximately 40\% wider and 40\% smaller domain in $y$-direction which show practically identical results to confirm that the width is large enough to neglect finite-size effects and small enough to suppress Plateau-Rayleigh instabilities. The total amount of fluid particles is set to $N = 4 \cdot 10^4$ per droplet. To control the temperature we use a dissipative particle dynamics (DPD) thermostat~\cite{Hoogerbrugge_1992, phillips_hoomd_dpd} in order to reproduce the correct hydrodynamics. In addition, we averaged every simulation setup over 50 trajectories to generate sufficient statistics.

To analyze the trajectories of the particles we first perform a projection along the $y$-axis so that we basically analyze a 2D-system (with axes $x$ and $z$ cf. Fig.~\ref{fig:snapshotMD}). 
A common procedure~\cite{Toxvaerd2007, weijs2011, kanduc2017} to extract the droplet shapes from these projections is to calculate the density field first and then determine the position of the liquid gas interface for different heights $z$. Thus, we calculate the position of the liquid vapor interface for every $z$-position by fitting the density with the function
\begin{equation}
\label{eq:md_fitting}
c_{z}(x) = \frac{1}{2}\bigl( c_{l} + c_{g} \bigr) - \frac{1}{2} \bigl(c_{l} - c_{g} \bigr)\tanh\Biggl(\frac{2(x-x_{\beta}(z))}{d_{\beta}} \Biggr).
\end{equation}
Here, $c(x)$, $c_{l}$ and $c_{g}$ are the particle density at position $x$, the density in the bulk and the density in the gas phase respectively. $x_{\beta}(z)$ gives the position of the liquid vapor
transition by using a crossing criterion. The parameter $d_{\beta}$ determines the width of the liquid-vapor interface. The same procedure can be applied for the perpendicular direction parallel to the $z$-axis to compute the liquid gas interface position $z_{\beta}$ for a given $x$ position.
 For our analysis we used the values of $x_{\beta}(z)$ for $z < h/2$ with a bin size of $1~\sigma$ to average out layering effects and the values of $z_{\beta}(x)$ for $z > h/2$ where we use a bin size of $0.1~\sigma$. Here, $h$ is the height of the droplet. Both methods can be consistently combined (cf. details in the Supplemental Material) and help to resolve the droplet peaks better, because all relevant fits are along lines with a significant share of particles in the liquid phase.
 Finally, we have chosen the height of the upper layer of the substrate to be at $z = -0.5 \sigma$ since the particles have a diameter of $\sigma$, i.e. the top of the substrate particle is at $z=0$.

\subsection{Thin film equation theory (TFE)}
The mesoscopic continuum model employed in this paper is based on the thin-film or lubrication approximation for the Navier-Stokes equation~\cite{OrDB1997rmp}. The lubrication approximation is given by an evolution equation of the local height $h=h\qty(x,y,t)$, which can be written in a gradient dynamics form as~\cite{MITLIN1993491,Thie2010}
\begin{equation}
\partial_t h = \nabla \cdot\qty[M(h)\nabla \fdv{\freeenergy}{h}] \label{eq:tfe_gradient_form}
\end{equation}
with the mobility $M(h)$ and the free energy functional $\freeenergy=\freeenergy\qty[h]$. Here, we assume no-slip boundary conditions at the substrate leading to a mobility of $M(h)=h^3/(3\eta)$ with the dynamic viscosity $\eta$~\cite{OrDB1997rmp}. Several slip regimes can be accounted for by different choices for $M(h)$~\cite{FJMWW-PRL-05}. For a discussion of the influence of the mobility on the dynamics cf. \cite{HLHT2015w,munch2005lubrication}.
The generalized pressure $P = \fdv{\freeenergy}{h}$ is given by
\begin{equation}
P(h,x,t) = -\gamma \Delta h - \Pi(h,t)\label{eq:pressure}
\end{equation}  
with the surface tension $\gamma$ and the disjoining pressure $\Pi(h,t)$. The latter is chosen as
\begin{equation}
\Pi(h,t) = \qty(\frac{B}{h^6}-\frac{A}{h^3})\qty(1+\rho(t)) \label{eq:disjoining}.
\end{equation}
Here, $A$ and $B$ denote the interaction strengths of long and short ranging forces respectively, where $A$ is directly connected to the Hamaker constant $H$ by $A=H/6\pi$. A different choice of the disjoining pressure is possible, see e.g.~\cite{deGennes-RMP-85,MITLIN1993491,OrDB1997rmp} for details. Even though the disjoining pressure determines mainly equilibrium properties an effect on dynamics is possible, because non-equilibrium transient states are inherently involved in dynamics.
In the present paper, the disjoining pressure in Eq.\,\eqref{eq:disjoining} is modulated by the parameter $\rho=\rho\qty(t)$ to model switchable substrates, where $\rho$ is a parameter corresponding to the wettability contrast. The use of the no-slip boundary condition at the substrate leads to a logarithmic energy dissipation at the contact line~\cite{BEIM2009romp}. This singularity can be resolved by introducing a precursor film $h=h_p$~\cite{BEIM2009romp, OrDB1997rmp}, which is also present on macroscopically \textquote{dry} parts of the substrate. Alternate ways to resolve the singularity problem at the contact line are presented by \citeauthor{BEIM2009romp}~\cite{BEIM2009romp}. Note that a temporal modulation of the disjoining pressure as shown in Eq.\,\eqref{eq:disjoining} does not change the precursor film height~\cite{HLHT2015w, TBHT2017tjocp}. 
 
In the following we employ the non-dimensionalized form of Eq.\,\eqref{eq:tfe_gradient_form}, where $h$, $x$ and $t$ are scaled in such a way, that $3\eta$, $\gamma$, $A$ and $B$ are incorporated in the corresponding scaling (A detailed derivation of the nondimensionalization is presented in the Supplemental Material). This leads to the evolution equation

\begin{equation}
\partial_t h = \nabla \cdot\qty{h^3\nabla\qty(-\Delta h -\frac{5}{3}\Theta_{\text{eq}}^2\chi^2\qty(\frac{\chi^3}{h^6}-\frac{1}{h^3})\qty[1+\rho(t)] )} \label{eq:tfe}
\end{equation}
with the equilibrium contact angle $\Theta_{\text{eq}}$ and the parameter $\chi =h_p/h_0$, where $h_0$ is the spatial scale. For further analysis we subtract the precursor film height $h_p$ from the film height $h$ in Eq.\,\eqref{eq:tfe}. Additionally, we keep $\Theta_{\text{eq}}=\sqrt{\frac{3}{5}}$ so that the effective equilibrium contact angle $\tilde{\Theta}_{\text{eq}}$  is determined by the parameter $\rho$ as
\begin{equation}
\tilde{\Theta}_{\text{eq}} = \Theta_{\text{eq}}\sqrt{1+ \rho\qty(t)}.
\end{equation} 
In the MD model there is no equivalent precursor film, so $\chi$ has to be chosen small. We use a value of $\chi=0.01$ in the following, because smaller values of $\chi$ do not change the contact region significantly and would led to increased computational efforts and numerical problems. Note that for different values of $\rho$ the ratio of the precursor film height $h_p$ to the maximum droplet height $h_{\text{max}}$ does change for constant volume.

It should be noted that temperature does not enter directly into the TFE model (cf. Eq.\,~\eqref{eq:tfe}). However, indirectly the temperature enters the TFE model through the surface tension $\gamma$, the viscosity $\eta$ and the particular shape of the disjoining pressure, which determines the wetting regime. In particular, the minimum of the interface potential (the integral of the disjoining pressure) is directly connected to the contact angle in the mesoscopic picture as a known disjoining pressure is sufficient to determine the equilibrium state~\cite{BEIM2009romp, HLHT2015w}. Direct temperature dependence can be incorporated in thin-film models, e.\,g. \citeauthor{davidovitch2005spreading} have shown that higher temperatures leads to faster spreading~\cite{davidovitch2005spreading}. In our case this would be incorporated via a lower viscosity.

The direct numerical simulations within the TFE model~\eqref{eq:tfe} are based on the finite element library oomph-lib~\cite{HeHa2006}. In contrast to the MD model, cylindrically shaped droplets can directly be simulated on a one-dimensional (1D) domain in the thin-film model, which reduces the spatial dimension of the problem by one and directly excludes any instabilities possibly occurring in the transversal $y$-direction.
In general, one can plug profiles from the MD model into the TFE model. As the MD model exhibits noise this requires small time steps in the simulation to reach a smooth profile. In some circumstances the time steps in the employed adaptive time stepping algorithm can get so small, that rounding errors of the machine can influence the results. To avoid such behavior, we apply a filter to the MD data before we start the simulation in the TFE model. Details on the applied LOWESS filter can be found in the Supplemental Material.  
\section{Results}
\subsection{How to compare scales}
\label{ssec:how_to_compare_scales}

In order to compare the scales in case of a static droplet, a mapping between $\eps{w}$ and $\rho$ responsible for the wettability in their respective models is necessary. For dynamic comparisons the time scales need to be mapped as well. In the first part of this section we describe the mapping in the static, equilibrium case and in the second part the mapping of time scales.

The relation of the liquid-solid interaction strength to the contact angle has been the focus of research for quite some time, e.\,g. \citeauthor{RefA_review}~\cite{RefA_review} came up with a qualitative theory on the base of a van der Waals model in 1981. This is closely related to the research on wetting transitions substantially advanced by \cite{RefB_review,RefC_review,RefD_review,RefE_review}. \citeauthor{RefB_review}~\cite{RefB_review} investigated the wetting transition within a lattice-gas model for different interaction strengths and interaction ranges. Within the framework of a systematic van der Waals theory (mean-field model) critical wetting can also be observed~\cite{RefC_review}. Such critical wetting is non-generic as shown first by \citeauthor{RefD_review}~\cite{RefD_review}. 

Despite the efforts in this field there is no way to compute the relation between the interaction strength $\eps{w}$ and the wettability parameter $\rho$ without doing involved numerics. Even if possible, theories often only promise qualitative agreements~\cite{RefA_review}, which is insufficient for our aim of a quantitative agreement in the dynamics of microscopic and mesoscopic models.

For even better agreement of the equilibrium droplet shapes one could try to extract the exact shape of the disjoining pressure $\Pi$. As mentioned above, the disjoining  potential can in principle be extracted from MD simulations, lattice density functional theory (DFT), and continuum DFT~\cite{TMTT2013tjocpb,RefI_review, RefJ_review,RefK_review}. However,the extraction of wetting potential is tricky and some open questions remain~\cite{RefK_review}. 

Here we choose a different approach and map the interaction strength $\eps{w}$ from the MD model to the parameter $\rho$ from the TFE model so that according to an appropriate criterion both models show the same droplet shape. Since the volume is fixed, the shape of the static droplet is characterized by just one parameter such as the height or the contact angle. As a consequence, $\rho$ is used for a reliable, yet, empirical static mapping. This allows us to take the next step of a quantitative comparison not only of statics but also of dynamics. 

\begin{figure}
\includegraphics[width=\columnwidth]{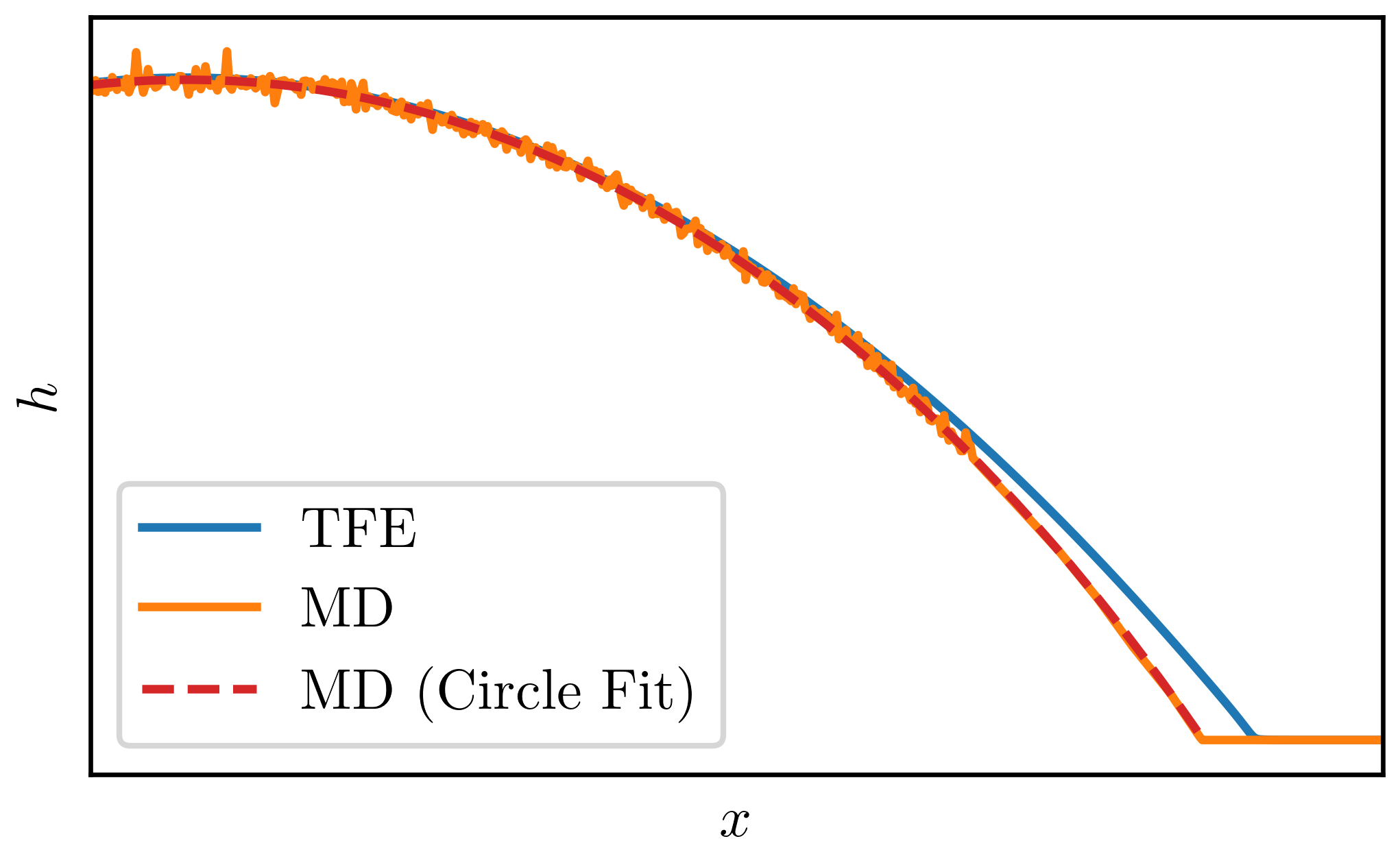}
\caption{Spatially rescaled height profiles of a static droplet from the MD and TFE model corresponding to  $\eps{w} = 0.762$ and a circle fit to the height profile from the MD model.}
\label{fig:contactregion}
\end{figure}

Contact angles are hard to define and measure consistently in microscopic and mesoscopic models. The definition of the contact angle can have a strong influence and at the nanoscale contact angle can depend on the droplet size~\cite{KuGa2021phys.rev.fluids}. Indeed, in the mesoscopic TFE model the contact region is not represented accurately due to the lubrication approximation~\cite{BEIM2009romp} and the contact line is hard to define due to the necessarily smooth transition to the precursor film. Consequently, it is not clear how results based on a contact angle mapping can be interpreted reliably. Figure\,\ref{fig:contactregion} shows two mapped equilibrium profiles based on our mapping described in the following to give an idea, how the lubrication approximation (blue curve) influences equilibrium droplet shapes in the contact region. 

\begin{figure}
\includegraphics[width=\columnwidth]{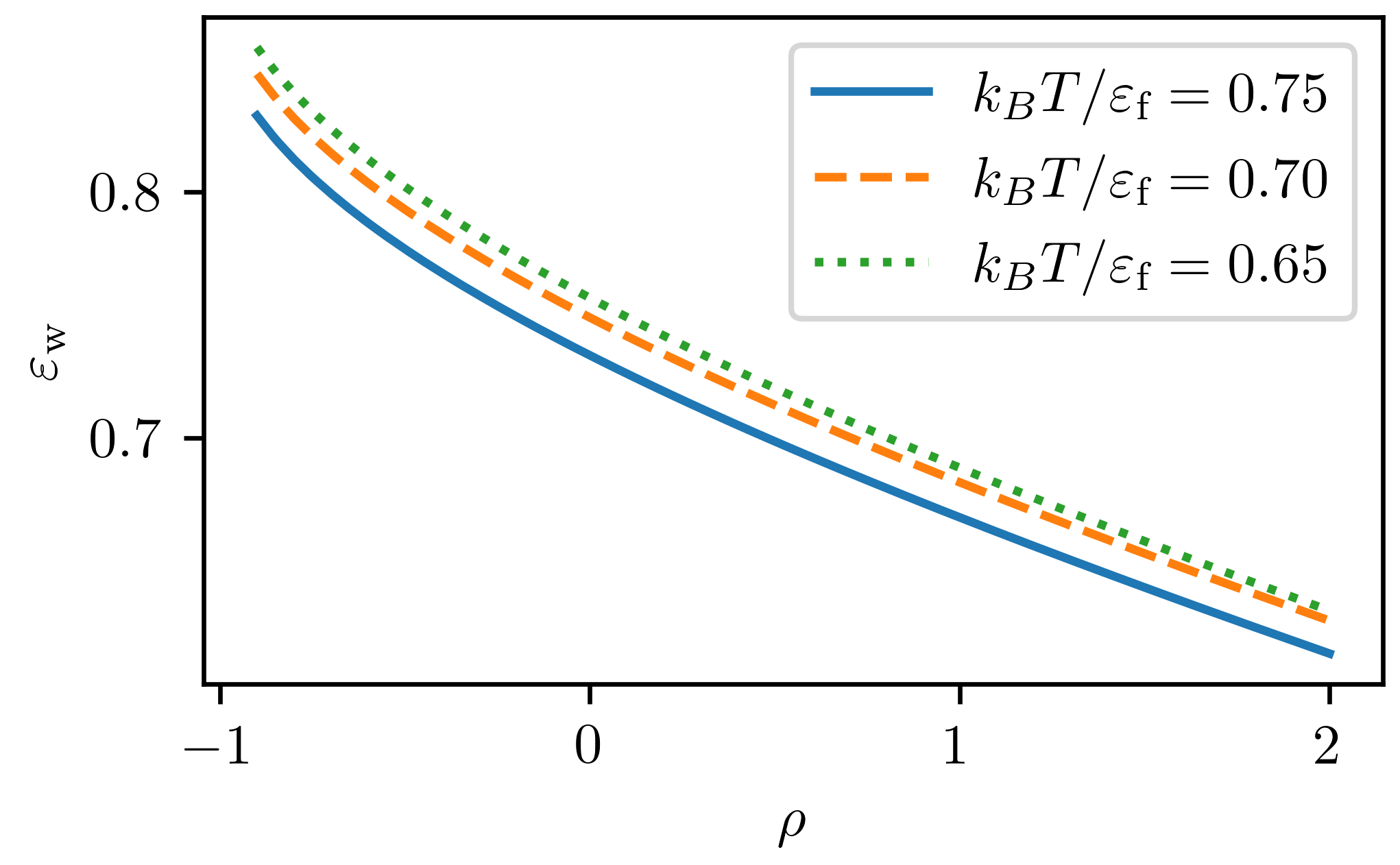}
\caption{Resulting mapping from the parameter $\rho$ modulating the disjoining pressure in the thin-film picture to the interaction strength between solid and liquid $\eps{w}$ in the MD model. The mapping is shown for multiple reduced temperatures. $k_B T/\epsilon=0.75$ is the value used for all the simulations within this manuscript.}
\label{fig:eps_vs_rho}
\end{figure}

Instead, we introduce \emph{the relative full width at half maximum} $\text{rFWHM}$, which is defined as the height of the droplet divided by its width at half of the height. This parameter is not sensitive to the droplet shape in the contact region, while being sensitive to the overall shape. The measure $\text{rFWHM}$ can be regarded as a computationally cheap way to estimate the curvature and thus is closely related to the contact angle. Another advantage of the $\text{rFWHM}$ is that one can analyze the temporal evolution in a straight-forward manner. If one measures the contact angle by fitting the droplet shape with a circular cap, the quality of the fit varies, because dynamic droplets do not necessarily have a circular cap shape and thus the error margin of the measured contact angle changes.

We computed the equilibrium droplet shapes in both models and measured rFWHM in dependence of $\eps{w}$ for the MD model and as a function of $\rho$ for the TFE model. 
The $\text{rFWHM}$ shows an approximately linear dependence of the interaction parameter squared $\eps{w}^2$, so that we employed a linear fit to avoid a computationally costly parameter sweep with a higher resolution in $\eps{w}$. The low computational cost in the TFE model makes a detailed parameter sweep possible, so that linear interpolation can be used to compute the rFWHM for arbitrary values of $\rho$ and vice versa. Combining both results yields a reliable, numerical mapping between $\eps{w}$ and $\rho$ as can be seen in Fig.\,\ref{fig:eps_vs_rho}. Additionally, Figure\,\ref{fig:eps_vs_rho} shows the mapping for different reduced temperatures. At higher temperature a defined interaction strength $\epsilon_w$ is mapped to a lower value of $\rho$, which corresponds to a lower contact angle. This explains how the disjoining pressure can capture the temperature dependence implicitly in the TFE model.

For the mapping of the spatial scales the $x$- and $z$-coordinates can be simply normalized with the maximum height of the droplet in the corresponding model at a certain wettability.
Here, the maximum height of a droplet on a surface with a wettability corresponding to $\eps{w} = 0.632$ in both models is used to scale the film height and the $x-$coordinates. In principle the scaling height can be chosen arbitrarily as long as it corresponds to the same height in both models, i.\,e. droplet height at a certain wettability.

A comparison of static droplet shapes from both models can be seen in Fig.~\ref{fig:contactregion}. Both profiles match well for heights of $h > 0.4\,h_{\text{max}}$. As anticipated, there is some deviation in the contact region due to the lubrication approximation in the TFE model. 

The mobility $M$ is particularly important for the dynamics, as it influences the pathway towards the minimum in the free energy~\cite{HLHT2015w}. In the context of hydrodynamic TFE, the mainly discussed expressions cases of no slip at the solid substrate ($M\sim h^3$), weak slip with slip length $l_s$ at the substrate ($M\sim h^2(h+l_s)$) and intermediate slip ($M\sim h^2$)~\cite{MuWW2005jem}. Furthermore, by interpreting very small $h$ rather as an adsorption normalized by a constant liquid bulk density than a film height, recently the cases of the transport via diffusion of the entire adsorbed film ($M\sim h$) as well as transport via diffusion of a surface layer on the deposit ($M\sim\mathrm{const}$) as in typical solid-on-solid models were discussed~\cite{THIELE2018487,YinPRE17}.
This implies in particular, a mobility function $M \sim h^3 +\mathrm{const}$ would automatically switch between diffusive and convective transport when going from the adsorption layer to a droplet or thick liquid film~\cite{YinPRE17}. 

Before it is even possible to assess, whether a mobility has been chosen correctly, an adequate rescaling of the time for the comparison of the dynamics between MD and TFE model is necessary. 

In order to achieve a rather generic approach we compute the deviation $\Delta = |K_{\text{MD}}-K_{\text{TFE}}|$ of a measure $K$ from the TFE profiles for every MD time step first which is defined as
\begin{equation}
  K = \underset{h>0.4 h_{\text{max}}}{\int}  \frac{x^{2} h}{(\int h~\dd x)^{2}} \dd x.
\end{equation}
An advantage of this choice of $K$ is that it could be applied to different use cases like a droplet adapting to a new surface wettability or the coalescence of two droplets as shown later. This is not possible by taking a measure as the rFWHM because it is only defined for one single droplet. Note that it is possible to calculate the rFWHM from $K$ in the case $K$ is calculated for a system with a single droplet with the numerically obtained formula (details are given in the Supplemental Material)
\begin{equation}
    \text{rFWHM} = \frac{1}{0.0614 + 10.92 K}.
\end{equation}
To exclude systematic errors originating from the lubrication approximation we only consider regions with $h > 0.4\,h_{\text{max}}$. 

Since we are interested especially in a mapping of the dynamics we furthermore use a version of $K_{\text{MD}}$ and $K_{\text{TF}}$ that is normalized between 0 and 1 in order to compensate for small differences of $K$ present in the static droplet profiles. Such small errors would propagate to the dynamic mapping. The normalized $K$ is defined as 
\begin{equation}
K_{norm} = \frac{K - K(t=0)}{K(t=\infty) - K(t=0)} \label{eq:K_norm}. 
\end{equation}

With the help of the computed deviations we can match every time step in the MD simulation to the time step of the TFE simulation with the least amount of deviation. This results in a mapping $t_{\text{MD}}\mapsto t_{\text{TFE}}$. The mapping in this direction is more convenient as the MD model is a first principle model and uses a constant time step in contrast to the TFE model.

In the following we apply the resulting mapping to two different cases to demonstrate its applicability. First, we investigate the mapping for a liquid ridge placed on a homogeneous, switchable substrate and secondly, we consider the mapping in the case of coalescence of two ridges. There we show that our mapping method can be applied universally and does not require additional simulations or measurements, where one had to worry about initial conditions influencing the measurement of a characteristic time.
 
\subsection{Single switch on a homogeneous substrate}\label{sec:switching}
As a first example for the mapping procedure presented in Sec.\,\ref{ssec:how_to_compare_scales} we applied it to the dynamics of a one-dimensional ridge on a switchable substrate. The procedure for the simulations is as follows: The ridge is equilibrated at either high or low wettability ($\eps{HW}=0.762$ and $\eps{LW}=0.632$, respectively), before the wettability is instantaneously switched to the other wettability at $t=0$. 

\begin{figure}
\centering
\includegraphics[width=\columnwidth]{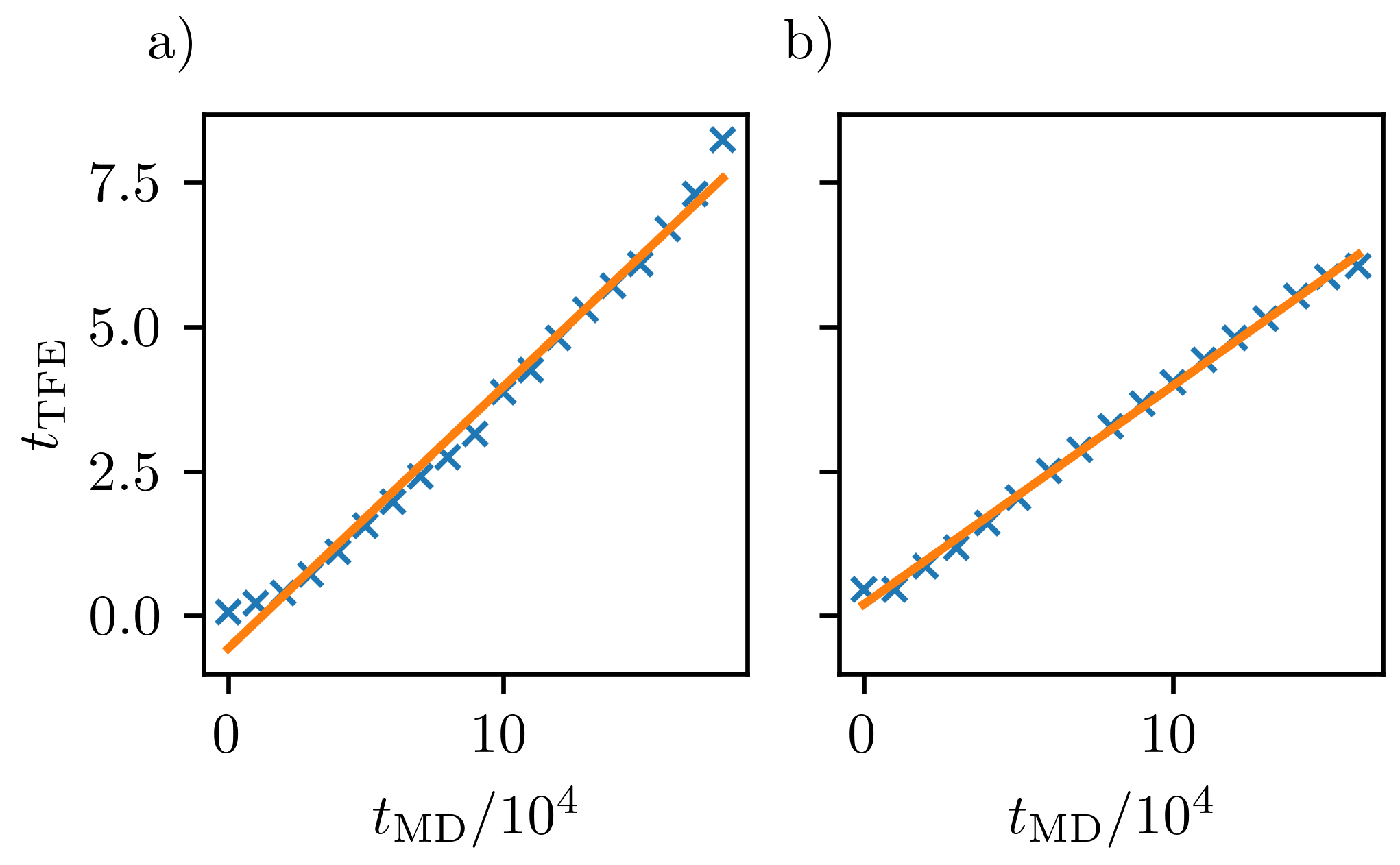}
\caption{Result of the mapping approach applied to a) the switching from $\eps{LW}$ to $\eps{HW}$ (blue data points and orange linear fit) and b) from $\eps{HW}$ to $\eps{LW}$. }\label{fig:compare_times}
\end{figure}

Figures \ref{fig:compare_times}~a,b show the results of the mapping approach for switching towards higher and lower wettability, respectively. In both cases the data points of the mapping hardly deviate from a linear fit.

Only the first two matched time steps deviate from the linear fit. This is possibly due to initial inertial effects in the MD model, which are not present in the TFE model~\cite{OrDB1997rmp, BEIM2009romp}. Consequently, a linear fit of the mapping does not go directly through the origin. As expected, deviations between the MD model and the TFE model exist in the contact region, corresponding to regions of the droplets with small heights. As a consequence the mapping is becoming worse for very high wettabilities because the contact region is more extended. A cut-off for large times needs to be introduced, because eventually changes in the droplet profile between time steps are dominated by noise in the MD model and matched times are not meaningful anymore. 

\begin{center}
 \begin{figure*}
\includegraphics[width=\columnwidth]{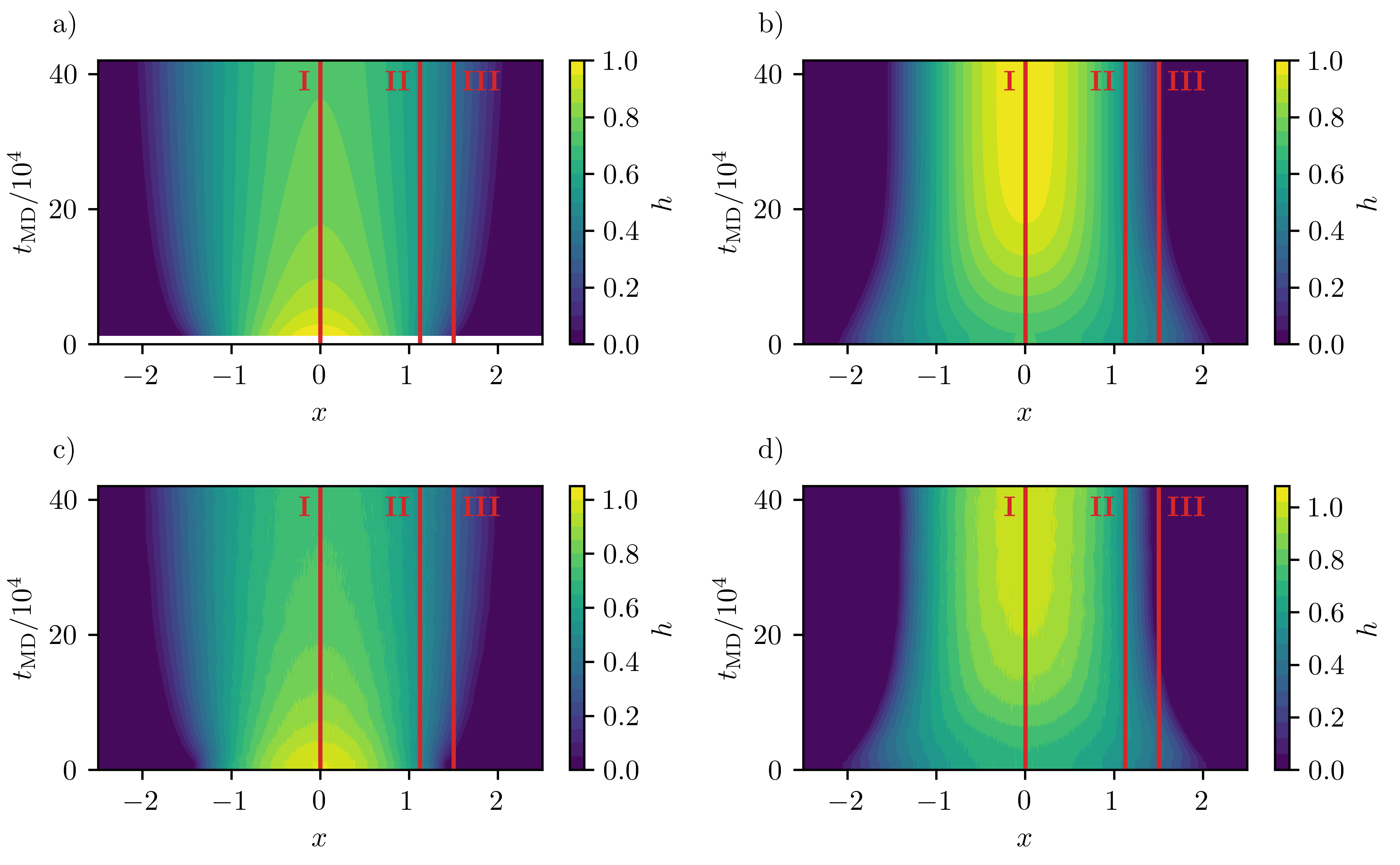}
\caption{Space-time plots showing the evolution of the height profile after a change in wettability for $\eps{HW}=0.762$ and $\eps{LW}=0.632$. a) and c) switch from low to high wettability in the TFE and the MD model, respectively. b) and d) show inverse switching direction in the TFE and the MD model, respectively. Red vertical lines labeled I, II and III indicate at which positions the height profile evolution is shown in Fig.\,\ref{fig:slices_spacetime}.}\label{fig:spacetimeplots}
\end{figure*}
\end{center}

\begin{figure*}
\centering
\includegraphics[width=\columnwidth]{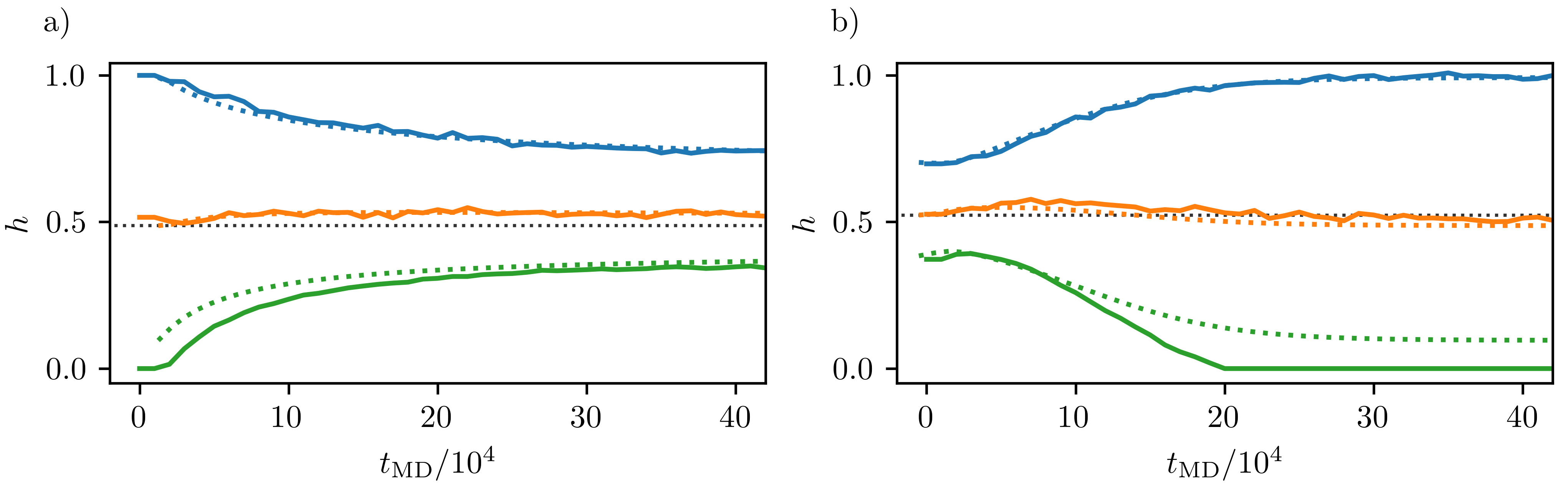}
\caption{Three characteristic height profiles along the marked lines I\,(blue), II\,(orange) and III(red) as shown in Figure\,\ref{fig:spacetimeplots}. The height profiles extracted from the MD model are presented with solid lines. The dotted lines are TFE results with the temporal scaling obtained using our mapping approach. The height $h$ is given relative to the maximum height. The black dotted line indicates the initial height in the TFE model for slice II to emphasize the non monotonous behavior for the height profile at this position. a) the wettability is switched from low to high, b) the wettability is switched in the opposite direction.}\label{fig:slices_spacetime}
\end{figure*}

The space-time representations in Fig.\,\ref{fig:spacetimeplots} show the evolution of the film profile $h(x,t)$ as the ridge adapts to the new wettability. The colormap indicates the film height. The results from both models look very similar in this representation. Besides the noise in the MD model the offset for the switch towards higher wettability accounts for the only general difference. This offset is a consequence of the previously described effect of the initial assimilation. To better grasp how good the models compare, the change of the height in time at three distinct points in space is shown in Fig.\,\ref{fig:slices_spacetime}. For the direct comparison between the models times from the TFE model are converted into MD units $t_{\mathrm{MD}}$.

For both switching directions the evolution for the height at points I and II is in a close agreement between the models.
The situation is however a bit different for point III. Only for maximum profile heights reached at this position the agreement is close. For smaller heights you can see differences. This behavior can easily be explained since already static ridges from both models do not match well in the region located close to the contact line (cf. Fig.\,\ref{fig:contactregion}). The reason is the underlying lubrication approximation in the TFE model, which results in higher film heights in the contact region compared to the MD model.

One can also find a position, where the height profile behaves non-monotonically. In particular, for point II the film height increases first and then decreases to its equilibrium value. This effect occurs for both switching directions, although it is more pronounced, when switching from high to low wettability. The dotted black lines in Fig.\,\ref{fig:slices_spacetime} are a guide to the eye to better see the non-monotonous behavior. 
In general, it is not surprising to find this non-monotonical behavior since this occurs also in the Gaussian solution of the heat equation or Fick's second law. However, finding such a distinct behavior in both models confirms that the parameters are chosen in such a way, that both models show consistent behavior, as this behavior could also be present at a different film height, a different position or a different point in time.

Table\,\ref{tab:ratios_from_compare_times} shows resulting time scale ratios (corresponding to the slope of the linear fit) for the switching process between different wettabilities. Note that the value of $R$ in general could depend on the initial and the final wettability. The data in Table\,\ref{tab:ratios_from_compare_times} does not indicate that there is a dependence on the initial wettability at least not within the error margin. The error margin is approximately 10\%, if you consider the influence of the cutoff for the fit. Notably, the time scale ratio $R$ shows a tendency to decrease if the wettability increases.  However, the resulting deviations are quite small (less than a factor of two in the considered range of wettabilities) and, furthermore, we observe a linear time mapping for all wettabilities. This shows that indeed the droplet evolution, seen in the MD simulations, is well reflected by the TFE.

\begin{table}
\caption{\label{tab:ratios_from_compare_times} Comparison of time scale ratios for different wettabilities and wettability differences computed with our mapping approach.}
\begin{ruledtabular}
\begin{tabular}{dddd}
\mbox{$\epsilon_{\text{LW}}$} & \mbox{$\epsilon_{\text{HW}}$} & \mbox{$R_{f_{\text{HW}\rightarrow \text{LW}}}$}& \mbox{$R_{f_{\text{LW}\rightarrow \text{HW}}}$}\\
\hline
0.632 & 0.671 & 2.98 \cdot 10^4 & 2.70 \cdot 10^4 \\ 0.671 & 0.707 & 2.95 \cdot 10^4 & 2.45 \cdot 10^4 \\ 0.707 & 0.742 & 2.13 \cdot 10^4 & 2.15 \cdot 10^4 \\ 0.632 & 0.762 & 2.64 \cdot 10^4 & 2.21 \cdot 10^4 \\ \end{tabular}
\end{ruledtabular}
\end{table}

\subsection{Coalescence of two ridges}\label{sec:coalescence_results}
As a second test case for the quality of the mapping procedure between the MD and TFE models we investigated the coalescence of two ridges on a homogeneous substrate. To get an input geometry a ridge on a surface with a wettability of $\eps{w} = 0.707$ is equilibrated in a MD simulation. 
A copy is then translated in $x$-direction in such a way, that it has a distance of $0.5~\sigma$ from the primary ridge at the contact line. To maintain the particle density we further extended the MD system (and thus the surface) in the $x$-direction by a factor of two. From this MD input geometry we calculated the positions for the liquid-vapor-transition of the two ridges and used this data as an input for the TFE simulations. 

\begin{figure}
\centering
\includegraphics[width=\columnwidth]{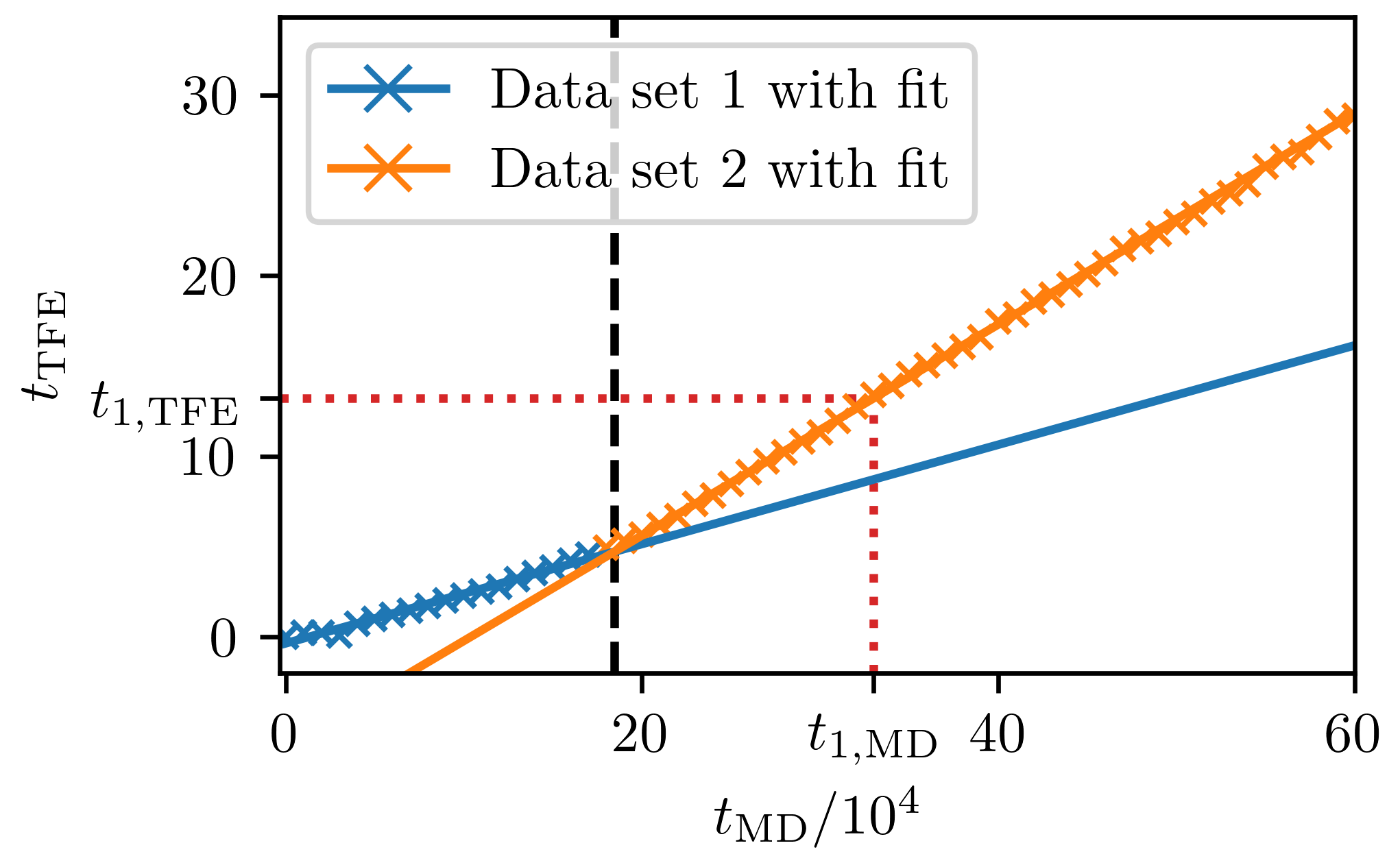}
\caption{Result of the temporal mapping for the coalescence of two ridges. Data points of the matched profiles are shown as crosses. Only the crossed data points in matching colors are considered for the corresponding fit shown as a solid line. Two distinct time scales can be observed, both fitted with a linear function. The slopes are $a_1 = 0.0264\pm 0.001$ and $a_2 = 0.05892 \pm 0.0003$. This corresponds to timescale ratios of $R_{f_{coal,1}}=3.79\cdot 10^4$ and $R_{f_{coal,2}}=1.70\cdot 10^4$ respectively. The black dashed line marks the intersection of the two linear fits at $t_{\text{MD}}=t_c=18.1\cdot 10^4$.}\label{fig:rescale_method_coalescence}
\end{figure}

The spatial rescaling from the single switch experiments can be employed without further issues. The result of our temporal mapping can be seen in Fig.\,\ref{fig:rescale_method_coalescence}. One can observe that in this cases the relation between between both models cannot be reasonably fitted with a single linear function. 
Instead, a piece-wise linear function is well suited to represent the data. In Fig.\,\ref{fig:rescale_method_coalescence} the colors of the linear fits match the data points used for fitting. The intersection of the linear functions is at $t_{\text{MD}}=t_c=18.1\cdot 10^4$. Figure\,\ref{fig:rescale_method_coalescence} visualizes, which times correspond based on our time scale mapping.

In particular, the time $t_c$ (black dashed line) corresponds to the time at which there is no longer a local minimum in the height profiles, i.e. the time at which the merging process of both droplets is completed. Plots of the height profiles corresponding to the matched times can be found in the Supplemental Material. After $t_{c}$ is reached, the droplet still contracts further. The time scale ratios are $R_{f_{coal,1}} = 3.64\times10^4$ in the first part and $R_{f_{coal,2}} = 1.71\times 10^4$  in the second part. 

Note that value of $R_{f_{coal,2}} = 1.71\times 10^4$ deviates not much from the value of $2.1\times 10^4$ which we observed for a single switch in Tab.\,\ref{tab:ratios_from_compare_times} for the final wettability $\eps{w}=0.707$. About the significant deviations for $R_{f_{coal,1}}$ we can only speculate. Possibly, the choice of the mobility in the TFE approach does not fully reflect the processes in the initial coalescence regime. Figure~\ref{fig:slices_coalescence} shows that the deviations between the MD and the TF profile are higher for $t\in\qty[0,t_c]$ especially for $x=0$. This could indicate that the path in phase space does not agree as good as in the case of a single switch and the disjoining pressure might need improvement.

It is also possible that an initial solution in the TFE model directly taken from the MD model encourages these deviations, because the differences in the contact region lead to an initial state in the TFE model, which is farther away from the equilibrium state than the initial state in the MD model. Consequently, the evolution could be artificially accelerated until the droplets have merged. Therefore, we performed additional simulations with initial droplet profiles taken from droplets equilibrated within the TFE model at the same peak distance. However, we basically obtained the same results. 

\begin{figure}
\centering
\includegraphics[width=\columnwidth]{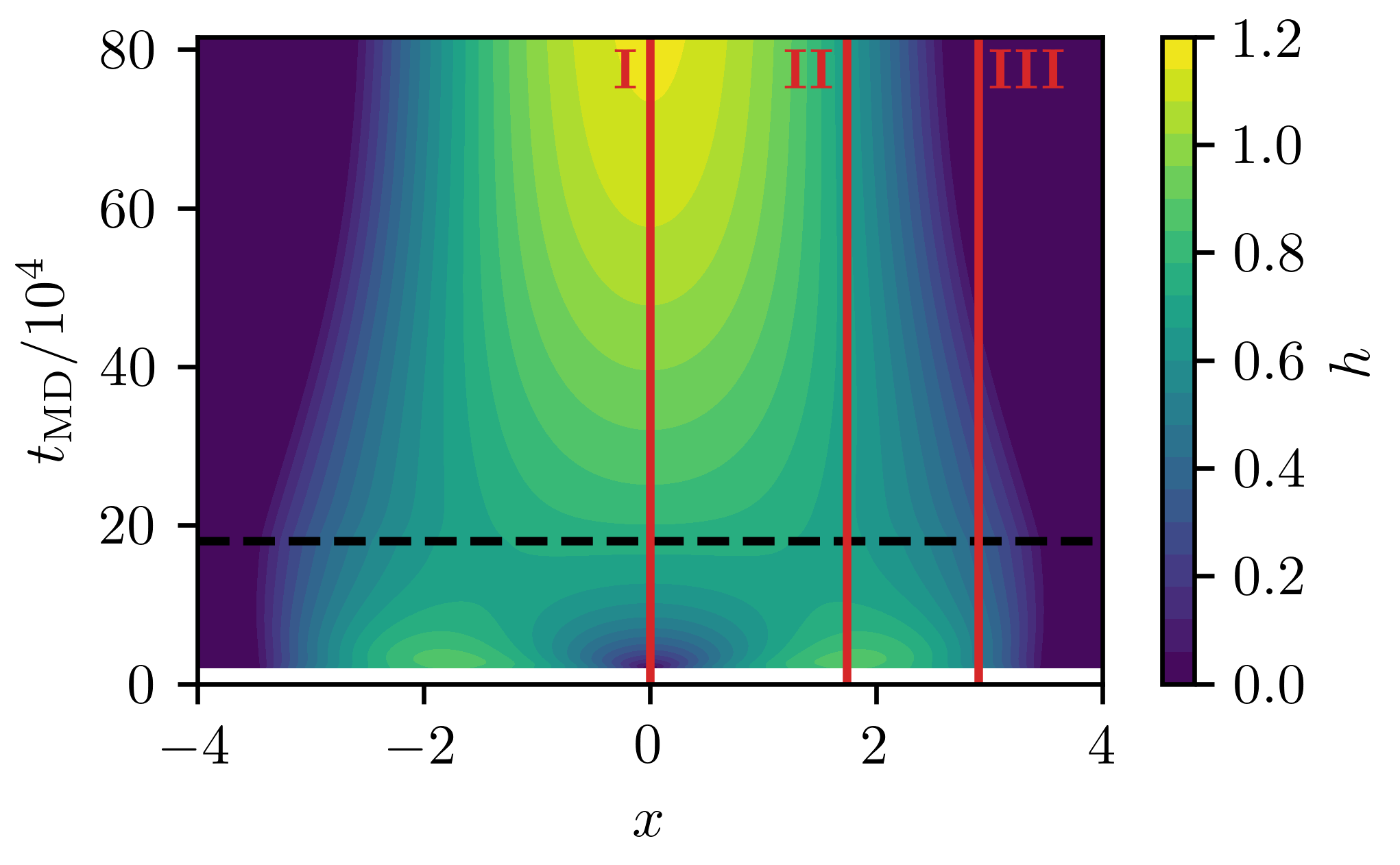}
\caption{Space-time plots showing the coalescence of two ridges in the TFE model. The black dashed line is where the TFE time scale is splitted and the linear fits in Fig.\,\ref{fig:rescale_method_coalescence} intersect. The height is scaled just as in Sec.\,\ref{sec:switching}. The space-time plot for the MD model is omitted,  because it is very similar to the TFE space-time plot. }\label{fig:spacetimeplots_coalescence}
\end{figure}

\begin{figure}
\centering
\includegraphics[width=\columnwidth]{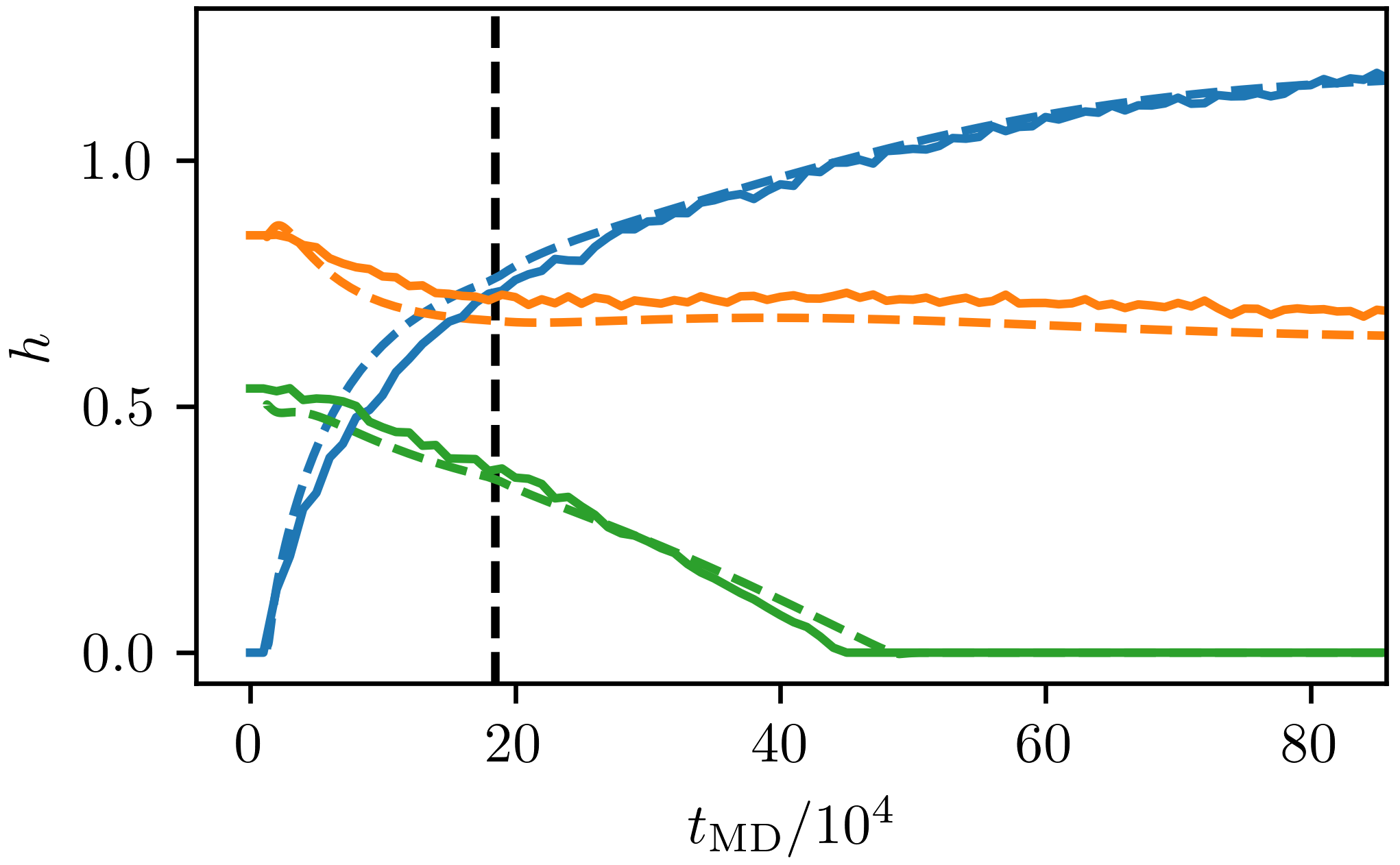}
\caption{Three characteristic height profiles along the marked lines I\,(blue), II\,(orange) and III(red) as shown in Fig.\,\ref{fig:spacetimeplots_coalescence}. The height profiles extracted from the MD model are represented with solid lines, whereas the height profiles from the TFE model are dashed. The scaling is the same as in Fig.\,\ref{fig:spacetimeplots_coalescence}. The dashed vertical line marks the time step, where the scaling factor is changed from $R_{f_{coal,1}}$ to $R_{f_{coal,2}}$.}\label{fig:slices_coalescence}
\end{figure}

Again, we can use our mapping approach to check the similarity of both approaches in the space-time plot; see  
Fig. \,\ref{fig:spacetimeplots_coalescence}. Furthermore, in Fig.\ref{fig:slices_coalescence} the evolution of the height profile at three distinct points is highlighted. The results match very well except for the anticipated difference at the contact region. Here we have used the strength of our mapping approach that we do not need a simple linear relation between time but can deal with arbitrary monotonous relations. 

\section{Conclusion}

In this paper we have presented a first step towards the quantitative comparison between the microscopic MD and the mesoscopic TFE model for the dynamics of liquid droplets. In a first step the parameters responsible for the substrate wettability in the respective models are mapped for equilibrium droplets via the rFWHM. Thereby, a mapping between the spatial quantities is achieved. 
In the second step, the general approach for the mapping of the time scales on the basis of matching the quantity $K$ which depends on the droplet shapes is applied. The applicability of the presented approach is shown thoroughly for two examples, a droplet on a homogeneous, switchable substrate and the coalescence of two droplets. We have demonstrated, that intermediate steps in the evolution are in good agreement. This implies, that the same path in phase space is taken in both models. This can be attributed to our reliable mapping of the parameters $\eps{w}$ and $\rho$ responsible for the wettability in their respective models.

Note that our mapping can always be applied, as long as one stays within the limits of the TFE models, i.\,e. low contact angle and concept of a precursor film. If one keeps the limits in mind, computational resources can be saved, because it enables one to accurately find the parameter range of interest in the TFE model and perform simulations on the mesoscopic scale, while still featuring the same quantitative non-equilibrium time evolution. The other direction is also possible, e.\,g. some interesting behavior is found within the TFE model and the MD model can provide microscopic insights of the dynamics. The presented approach is not specially tailored to the employed thin-film model. An application of the presented mapping approach for a continuum model, which allows for larger contact angles (e.\,g.  macroscopic boundary element method~\cite{GrSt2021softmatter}), is also possible.
Mapping to continuum models additionally promises fruitful results with the help of bifurcation analysis, which can help predict onsets of instabilities or explore large parameter spaces systematically along stable branches.

When studying in more detail the time mapping we found for the switching simulations that to a very good approximation the resulting mapping between both approaches does not depend on the initial wettability. This is indeed a very promising result, confirming the ability of the TFE approach to reproduce the microscopic results. Interestingly, two relevant deviations from a very simple mapping behavior could be observed. First, the factor $R$ between both time scales slightly depends on the final wettability. Second, in the coalescence we found a non-linear relation between time scales. These observations suggest that, e.g., the mobility or the disjoining pressure may need to be adapted for an even closer agreement. However, since in our approach the equilibrium properties are fully matched, the deviations are always of order unity. 

So far we have chosen a standard choice of mobility and disjoining pressure (cf. \cite{TBHT2017tjocp,HLHT2015w} among others) and we matched the most important characteristics of the disjoining pressure to the MD model, i.\,e. we made sure, that the minimum of the interface potential was correct, that the precursor film height was chosen adequately, and that the wetting regime was correctly mirrored by our disjoining pressure. In future our approach could be extended to a thin-film model with a fine-tuned disjoining pressure. Literature provides ample opportunities to extract the disjoining pressure directly from microscopic models, which enables comparisons between models in the static case~\cite{TMTT2013tjocpb,RefG_review,RefH_review,RefI_review,RefJ_review,RefK_review,RefL_review, RefN_review}. In contrast to the calculation of the disjoining pressure one need to consider dynamics to be able to extract information about the mobility from MD models.
Here we used a cubic mobility, which can be derived from no-slip boundary conditions. A precise mobility extracted from microscopic models promises to improve the TFE description of a MD model and is the subject of future work. 
Our procedure could serve as a measure to conclude, whether mobility and disjoining pressure were chosen adequately as you would expect a proportional relation of the time scales in the models. 

Furthermore, we would like to stress that in all cases the mapping of the times can be applied in a generic manner. This makes the mapping approach versatile and provides the ability to map time scales if characteristic times or other quantities necessary for a temporal mapping are not easily and reliably accessible.

  \label{sec:conclusion}

\begin{acknowledgments}
We wish to acknowledge the financial support of the DFG Schwerpunktprogramm SPP 2171.
\end{acknowledgments}

\section*{Data availability statement}
In the accompanying zenodo dataset (\url{http://doi.org/10.5281/zenodo.5147918})\cite{ZenodoDataset} we provide the simulation data, the code to reproduce the figures and the oomph-lib code to perform the simulations in the TFE model.

\newpage
\renewcommand{\appendixname}{Supplemental Material}
\appendix
\section{Droplet shapes}\label{app:droplet_shape}
\begin{figure}
\includegraphics[width=\columnwidth]{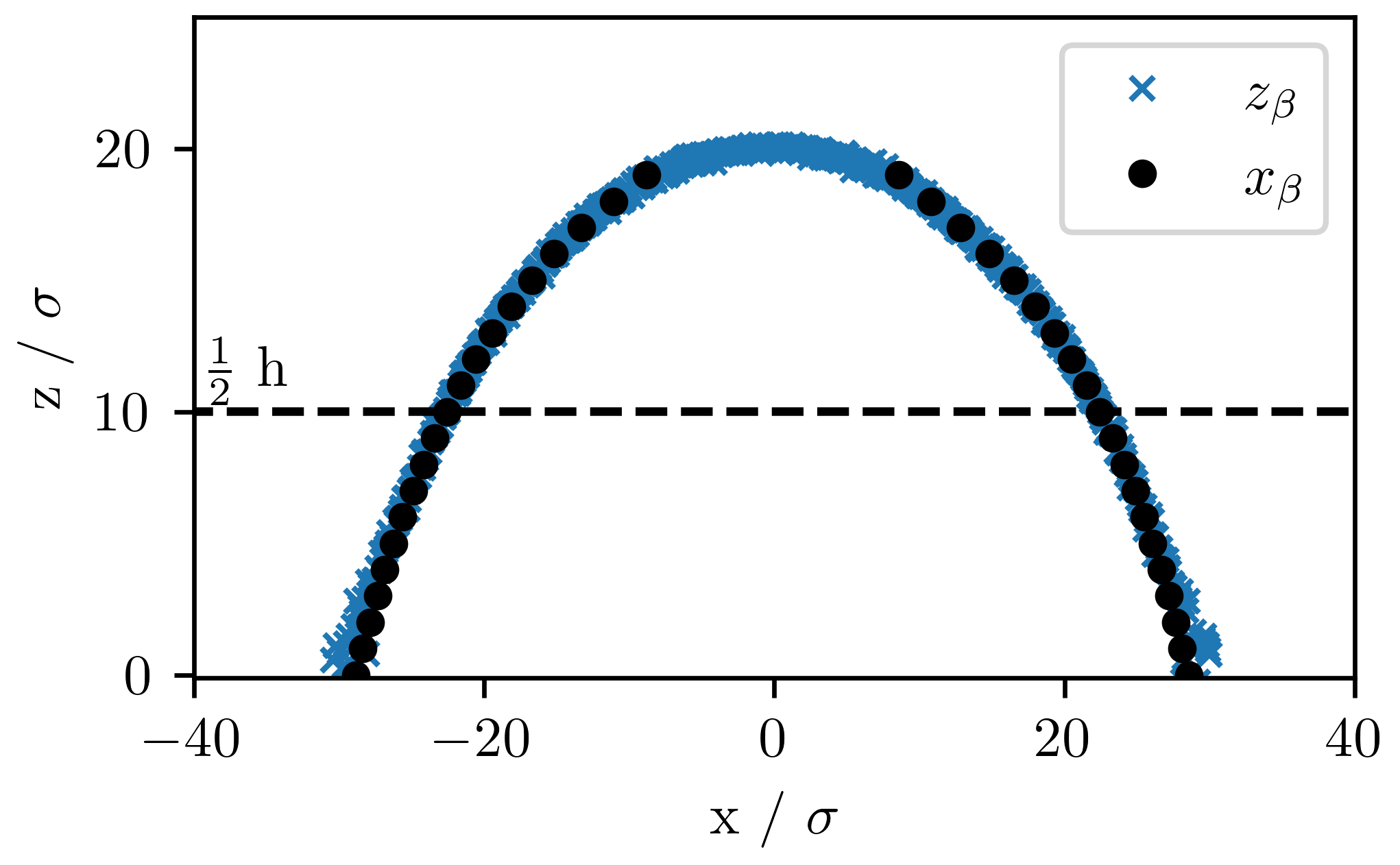}
\caption{Position of the liquid vapor interface obtained from fits parallel $x_{\beta}$ and perpendicular to the substrate $z_{\beta}$.}\label{fig:md_shape}
\end{figure}

Figure\,\ref{fig:md_shape} shows the position of the liquid vapor interface $x_{\beta}$ and $z_{\beta}$ calculated via fits of the density along the $x$- or $z$-direction respectively.
The fitting function
\begin{equation}
\label{eq:md_fitting_app}
c_{z}(x) = \frac{1}{2}\bigl( c_{l} + c_{g} \bigr) - \frac{1}{2} \bigl(c_{l} - c_{g} \bigr)\tanh\Biggl(\frac{2(x-x_{\beta}(z))}{d_{\beta}} \Biggr)
\end{equation}
is a step function, where $x_{\beta}$ is the position of the vapor interface. $c_l$ and $c_g$ are the liquid and gas density. The width of the interface is denoted as $d_{\beta}$. $z_{\beta}$ is determined analogously. For $z_{\beta}$ a bin width of 0.1~$\sigma$ is used. For $x_{\beta}$ a bin width of 1~$\sigma$ is used to average out layering effects. 

In Figure\,\ref{fig:md_shape} differences in the interface positions $x_{\beta}$ and $z_{\beta}$ are only significant at the droplet peak and in the contact region. This is because the fit quality is higher if $\rho_l$ and $\rho_g$ can be adequatly determined, which is only the case if a significant share along the fitted line is within the droplet. Consequently $z_{\beta}$ is less accurate in the contact region and $x_{\beta}$ is less accurate close to the droplet peak, which is why we combined both methods at $h = h_{\text{max}} / 2$.

\section{Nondimensionalization}\label{sec:nondimensionalization}
The thin-film equation with dimensioned variables reads
\begin{align}
    \partial_t h &= \nabla \left[ \frac{h^3}{3\eta}\, \nabla \frac{\delta F}{\delta h} \right]\\
    \frac{\delta F}{\delta h} &= -\gamma \Delta h - \qty(\frac{B}{h^6} - \frac{A}{h^3})\qty(1+\rho(t)).\label{eq:free_energy_variation}
\end{align}
For a given disjoining pressure $\Pi$ the precursor film height is defined as the minimum of the interface potential, which is defined as 
\begin{equation}
I =-\int \Pi dh = \qty[\frac{B}{5h^5}-\frac{A}{2h^2}]\qty(1+\rho(t)).
\end{equation}
The minimum of the interface potential is the zero of the disjoining pressure, which leads to
\begin{equation}
\frac{A}{h_p^6}-\frac{B}{h_p^3}=0 \quad \Rightarrow \quad h_p = \qty(\frac{B}{A})^{\frac{1}{3}} \label{eq:precursor_definition}
\end{equation}
In the context of the lubrication approximation the relation
\begin{equation}
\Theta_{\text{eq}}=\sqrt{\frac{-2I\qty(h_p)}{\gamma}}\label{eq:theta_eq_app}
\end{equation}
holds, which defines the equilibrium contact angle $\Theta_{\text{eq}}$ for a given interface potential $I$, surface tension $\gamma$ and precursor film height $h_p$. Combining Eq.\,\eqref{eq:precursor_definition} and Eq.\,\eqref{eq:theta_eq_app} yields the relations
\begin{equation}
A = \frac{5}{3}\gamma \Theta_{\text{eq}}^2 h_p^2 \quad\text{and}\quad B = \frac{5}{3}\gamma \Theta_{\text{eq}}^2 h_p^5
\end{equation}
for the constants $A$ and $B$. Plugging these into Eq.\,\eqref{eq:free_energy_variation} gives
\begin{equation}
\frac{\delta F}{\delta h} = -\gamma \Delta h - \frac{5}{3}\gamma \Theta_{\text{eq}}^2 h_p^2\left(\frac{h_p^3}{h^6} - \frac{1}{h^3}\right)\qty(1+\rho(t)).
\end{equation}
Then we introduce the scales
\begin{align}
    x \to x_0 \cdot \tilde x,~~~t \to t_0 \cdot \tilde t,~~~h \to h_0 \cdot \tilde h,~~~\frac{\delta F}{\delta h} \to F_0 \cdot \tilde {\frac{\delta F}{\delta h}}
\end{align}
and insert them into the dimensioned thin-film equation. After dropping the tildes we get 
\begin{align}
    \partial_t h &= \nabla \left[ \frac{h_0^2 t_0 F_0}{3\eta x_0^2}\, h^3 \nabla \frac{\delta F}{\delta h} \right]\\
    \frac{\delta F}{\delta h} &= -\frac{\gamma h_0}{x_0^2F_0}\Delta h - \frac{5\gamma \Theta_{\text{eq}}^2 h_p^2}{3F_0 h_0^3}\left(\frac{h_p^3}{h_0^3} \frac{1}{h^6} - \frac{1}{h^3}\right)\qty(1+\rho(t_0 t)).
\end{align}
The scales $x_0$, $h_0$, $t_0$ and $F_0$ can be chosen freely. Here, the parameter $h_0$ is used as a reference height, which can be an experimentally measured droplet size. We can then simplify the thin-film equation by solving
\begin{align}
\frac{h_0^2 t_0 F_0}{3\eta x_0^2} = 1 \\
\frac{\gamma h_0}{x_0^2F_0} = 1 \\
\frac{5\gamma \Theta_{\text{eq}}^2 h_p^2}{3F_0 h_0^3} = \frac{5}{3}\Theta_{\text{eq}}^2\chi^2,
\end{align}
where he have introduced $\chi=h_p/h_0$. Solving this system of equations for the scales eliminates most of the prefactors. Keeping the term $\frac{5}{3}\Theta_{\text{eq}}$ ensures that all the spatial dimensions are scaled similarly. The solution is
\begin{align}
    x_0 = h_0,~~~~
    t_0 = \frac{3\eta h_0}{\gamma},~~~~
    F_0 = \frac{\gamma}{h_0}.
\end{align}
With these scales the thin-film equation can finally be written as
\begin{align}
\partial_t h &= \nabla \left[ \, h^3 \, \nabla \frac{\delta F}{\delta h} \, \right]\\
\frac{\delta F}{\delta h} &= -\Delta h -\frac{5}{3}\Theta_{\text{eq}}^2\chi^2\qty( \frac{\chi^3}{h^6} - \frac{1}{h^3})\qty(1+\rho(t)). \label{eq:with_chi_and_theta}
\end{align}
In this form the temporal modulation of the disjoining pressure $\rho(t)$ is expressed in the dimensionless units. A more detailed derivation of the non-dimensionalization is presented by \citeauthor{Engelnkemper2017} \cite{Engelnkemper2017}.

\section{Details on the measure used for the dynamic mapping}

\begin{figure}
\centering
\includegraphics[width=\columnwidth]{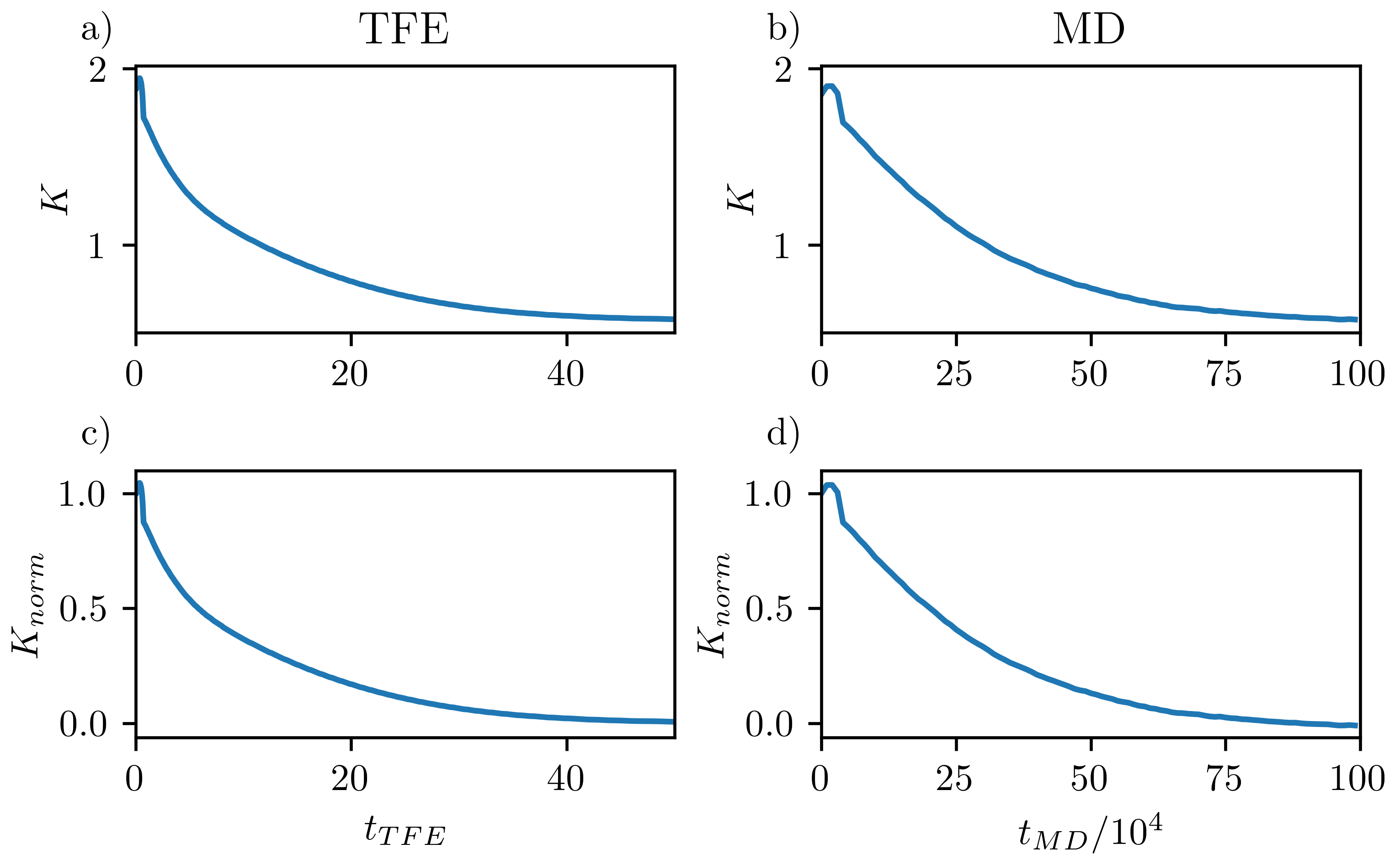}
\caption{Evolution of $K$ and $K_{norm}$ for the TFE (cf. a), c)) and the MD model (cf. b), d)).}
\label{fig:K_vs_K_normalised}
\end{figure}

We compute the measure $K$ according to
\begin{equation}
  K = \underset{h>0.4 h_{\text{max}}}{\int}  \frac{x^{2} h}{(\int h~\dd x)^{2}} \dd x. 
\end{equation}

Such values for $K$ differ slightly for equilibrium profiles between MD and TFE model (compare Fig. \ref{fig:K_vs_K_normalised} a) and b)) . The small errors would propagate to the dynamic mapping. Hence, we introduced a normalised $K$ defined as
\begin{equation}
K_{norm} = \frac{K - K(t=0)}{K(t=\infty) - K(t=0)} \label{eq:K_norm_SI}. 
\end{equation}
To reduce effects of noise in the MD model, we compute the value $K(t=\infty)$ as an average over a time span of $\Delta t=20\cdot 10^4$. In Figure \ref{fig:K_vs_K_normalised} $K$ and $K_{norm}$ are shown in the case of coalescence discussed in the main manuscript. Most importantly $K$ and $K_{norm}$ behave monotonically except for a brief time initially. For a single instantaneous switch the behavior is monotonous without exception.

\begin{figure}
\centering
\includegraphics[width=\columnwidth]{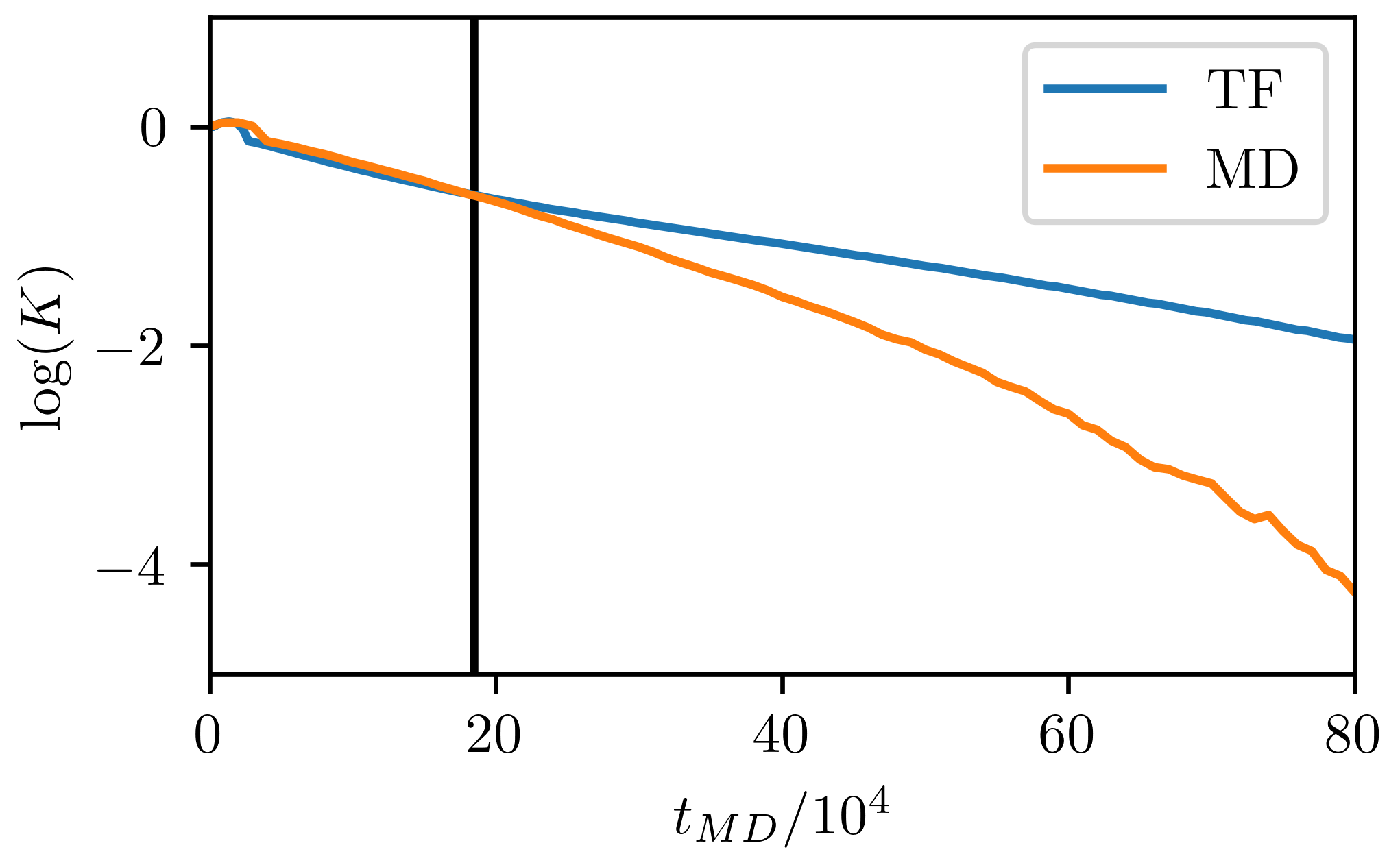}
\caption{Evolution of $K_{norm}$ on a logarithmic scale for the coalescence simulations. $t_{TFE}$ was scaled by the initial slope of the matched times. The solid black lines corresponds to the intersection of the two linear fits of the matched time.}
\label{fig:K_logscale}
\end{figure}

Figure \ref{fig:K_logscale} shows the evolution of $K_{norm}$ on a logarithmic scale. The logarithmic scale is able to visualize that a single linear fit of the mapped times is not sufficient. This is not obvious only looking at Fig. \ref{fig:K_vs_K_normalised}. 

\begin{figure}
\centering
\includegraphics[width=\columnwidth]{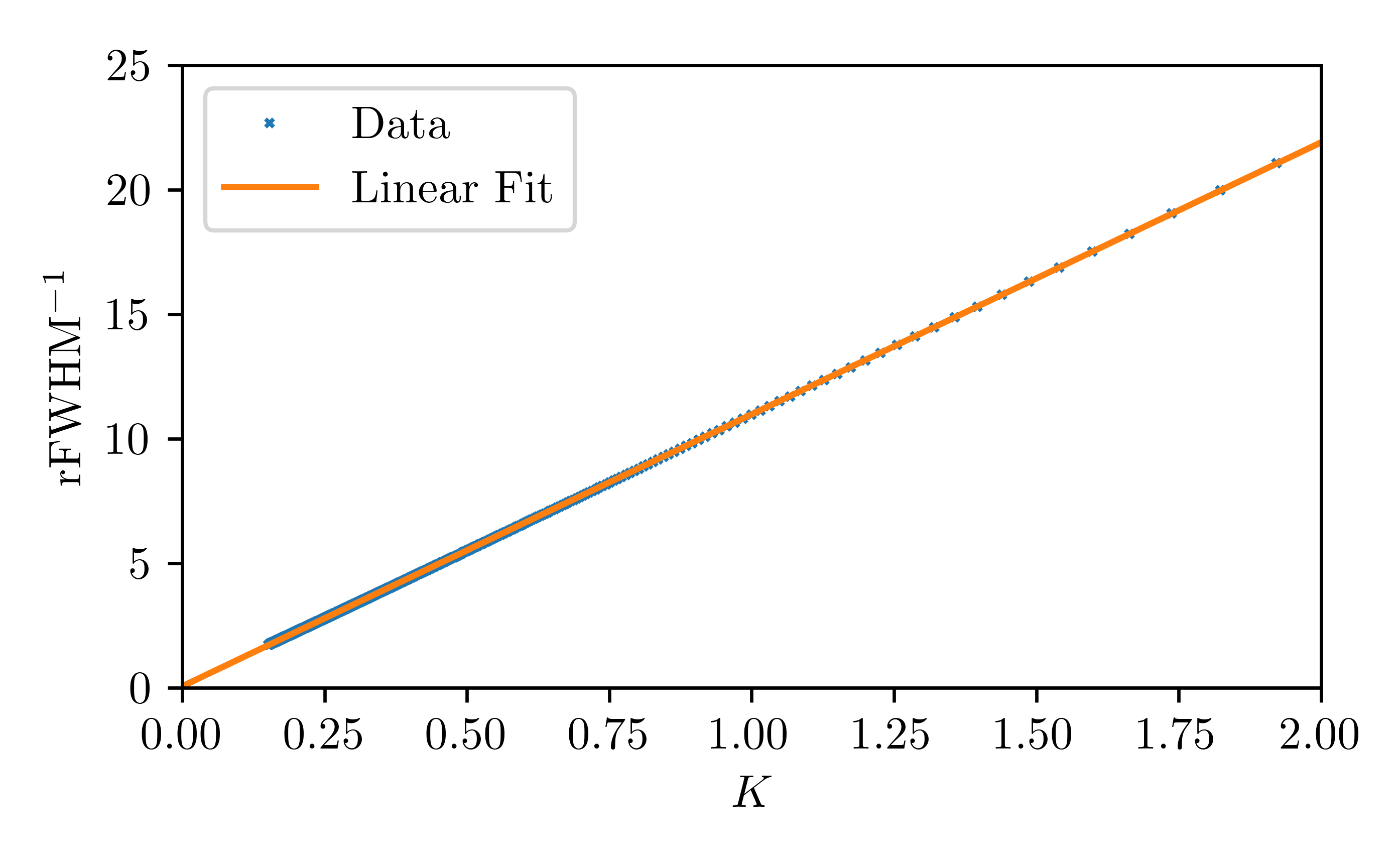}
\caption{$K$ vs. rFWHM$^-1$ under the assumption of circularly shaped droplets and a linear fit to the data. The fit parameters are shown in Eq.~\ref{eq:rFWHM_vs_K_app}.}
\label{fig:rFWHM_vs_K}
\end{figure}

The relation between $K$ and rFWHM
\begin{equation}
    \text{rFWHM} = \frac{1}{0.0614 + 10.92 K}\label{eq:rFWHM_vs_K_app}
\end{equation}
is a result of a numerical computation shown in Fig. \ref{fig:rFWHM_vs_K} under the assumption of circular shape. 

\section{Filter noise from MD profiles}\label{app:filter_noise}
\begin{figure}
\includegraphics[width=\columnwidth]{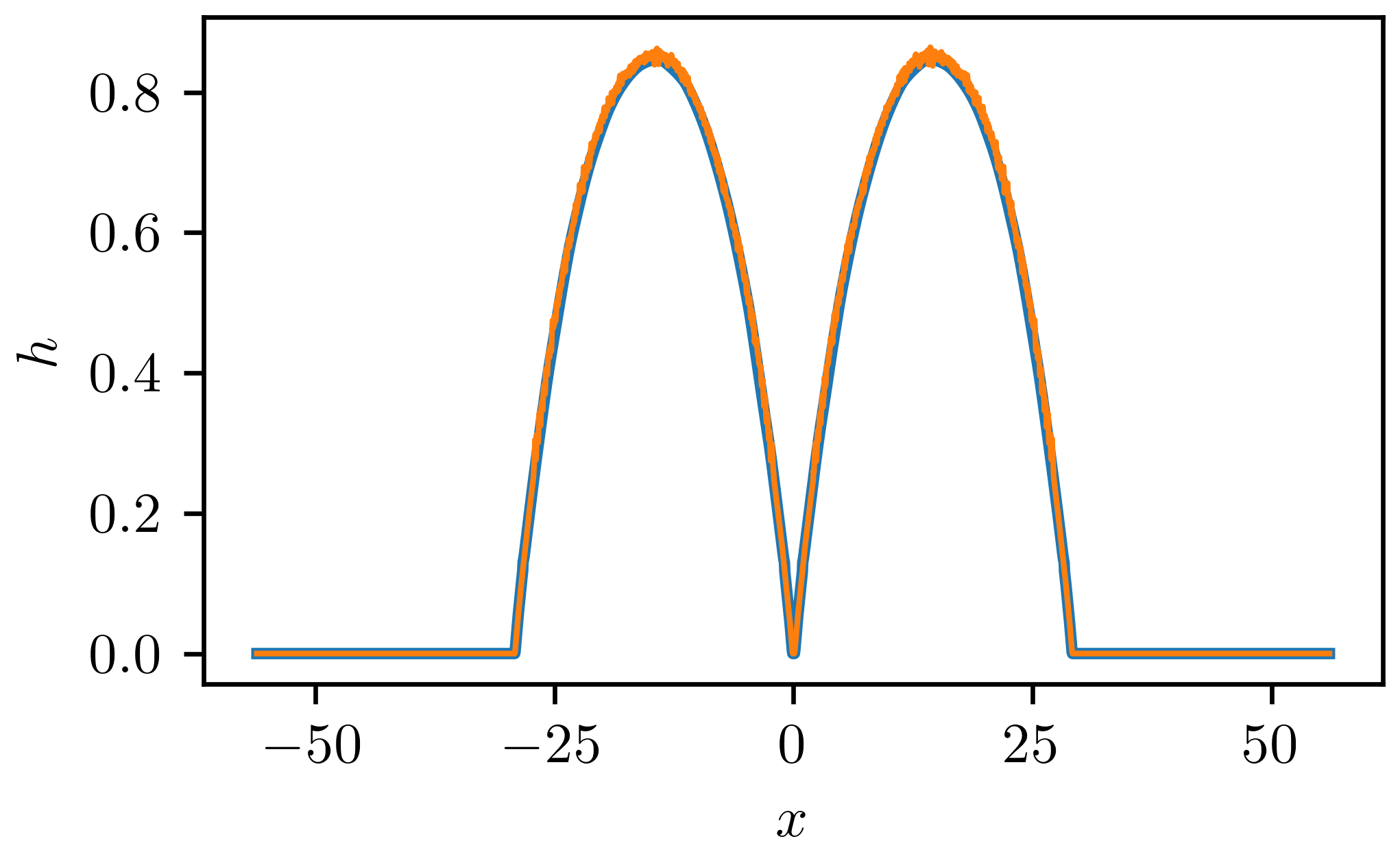}
\caption{Comparison of the height profile directly extracted from the MD model and the height profile after a LOWESS filter was applied to the parts of the MD data, where $h>20h_p$. }\label{fig:filter_vs_no_filter}
\end{figure}

Before a height profile from the MD model is plugged into the TFE model the noise is reduced on the height profile by averaging over 50 runs. Some noise still persists, which can lead to very small time steps in the TFE model possibly leading to inaccurate results due to rounding errors caused by the finite computational accuracy. Figure\,\ref{fig:filter_vs_no_filter} shows the difference between the filtered and the unfiltered data. We applied the LOWESS (Locally Weighted Scatterplot Smoothing) filter from the python package statsmodel \cite{SePe2010} to the height profile and only exchanged the data points with a height above 20 times the precursor film height. Otherwise the contact region would have been altered too much and noise in the contact region is low compared to the droplet peaks.

\section{Matched profiles when comparing time scales}\label{app:matched_profiles}

\begin{figure}  
\begin{tabular}{l}
a)\\
\includegraphics[width=0.95\columnwidth]{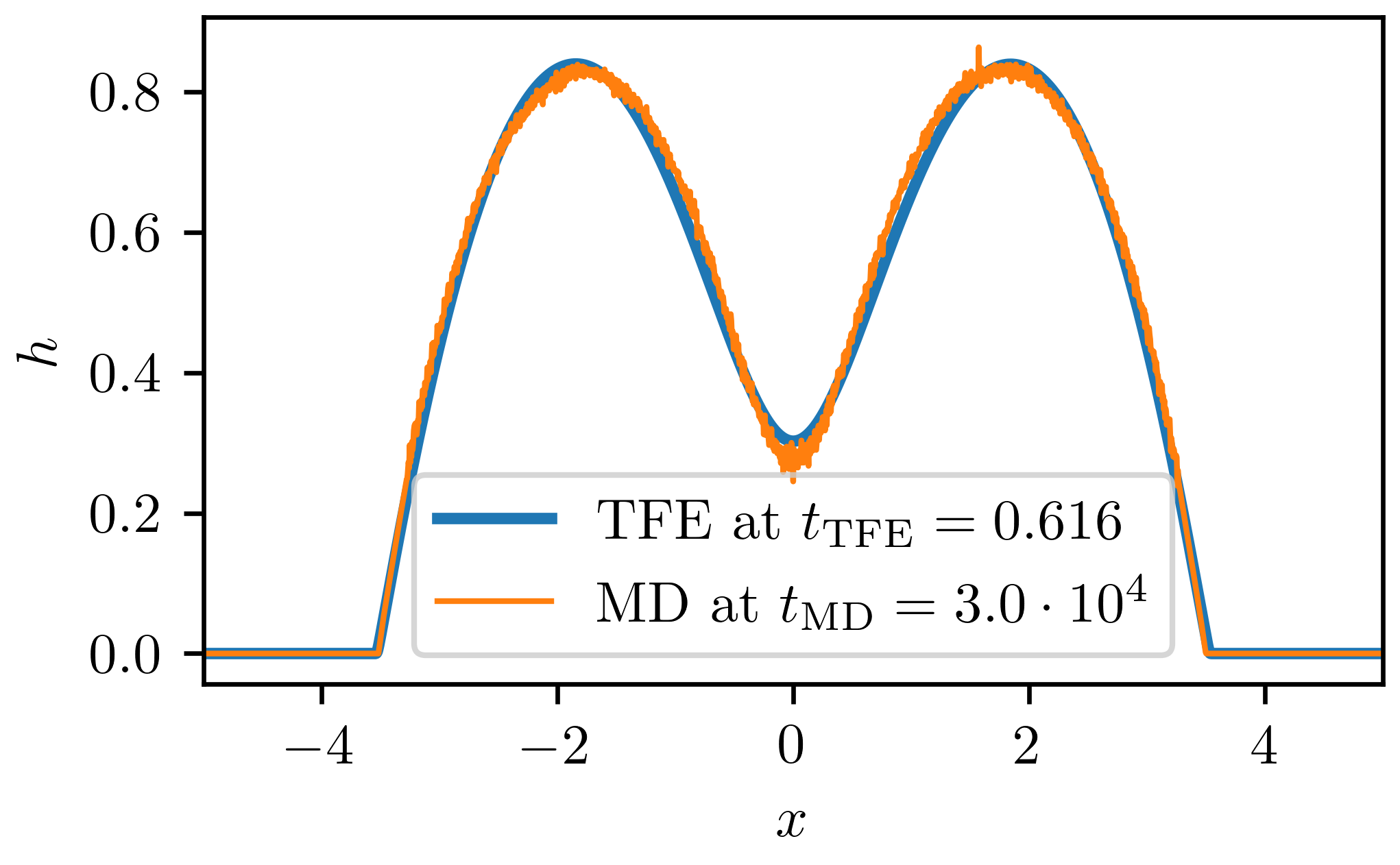}\\
b)\\
\includegraphics[width=0.95\columnwidth]{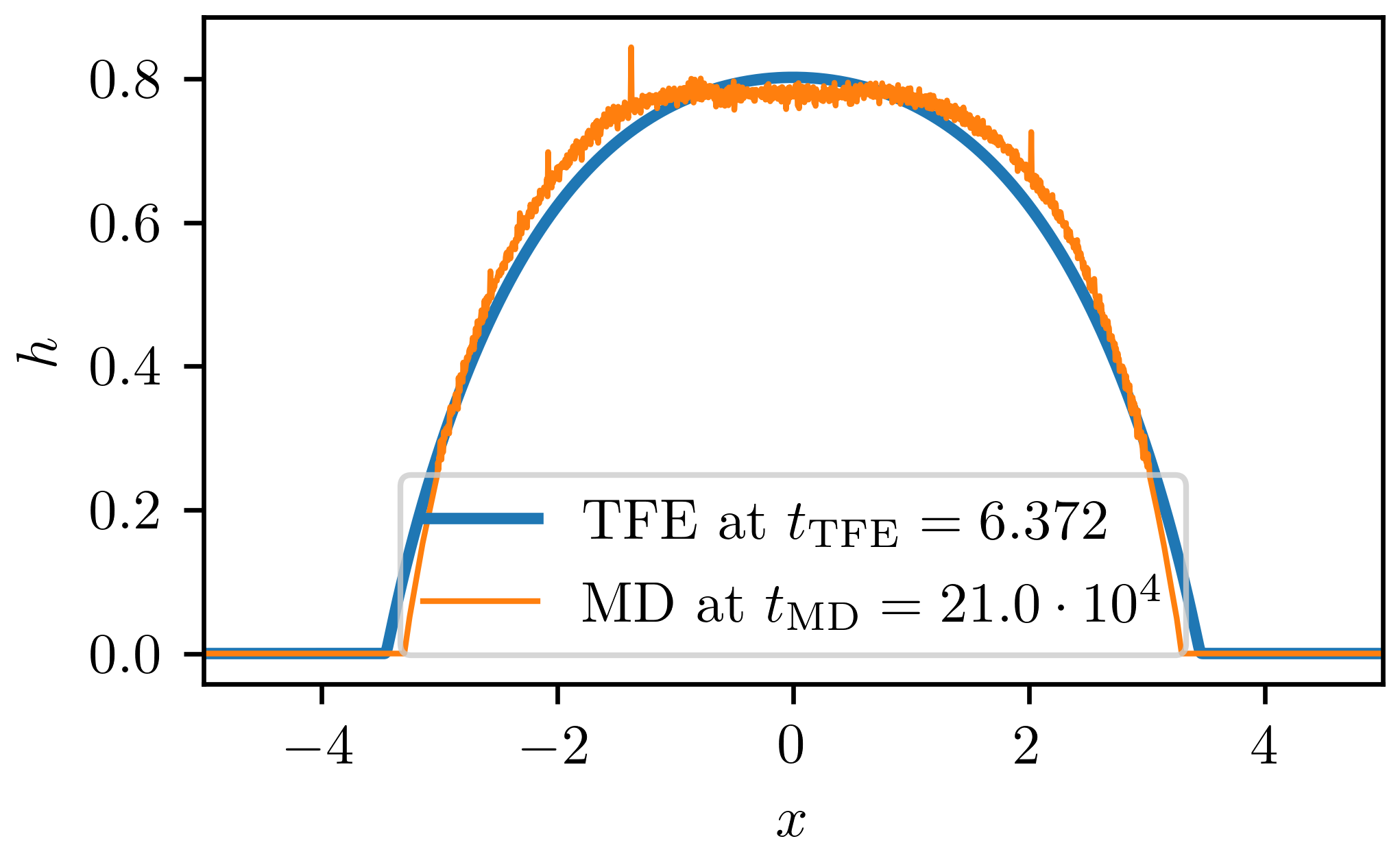}\\
\end{tabular}
\caption{Two examples of a height profile from MD simulations matched to a height profile from the TFE model taken from the coalescence simulation shown in Fig.\,\ref{fig:spacetimeplots_coalescence} of the main manuscript. }\label{fig:compare_TFE_MD_example}
\end{figure}
Figure\,\ref{fig:compare_TFE_MD_example} shows two matched height profiles according to our mapping procedure for the coalescence simulations investigated in Sec.\,\ref{sec:coalescence_results} of the main manuscript. The deviation at the contact region is expected as a consequence of the lubrication approximation. The matched profiles in b) correspond to times close to the critical time $t_c$, where both linear fits intersect in Fig.\,\ref{fig:rescale_method_coalescence} of the main manuscript.

\end{document}